\documentclass{aa}  

\usepackage{graphicx}
\usepackage{multirow}
\usepackage{txfonts}
\usepackage{hyperref}
\hypersetup{
    colorlinks=true,
    citecolor=blue,
    filecolor=magenta
}

\begin{document}

   \title{AMICO-COSMOS galaxy cluster and group catalogue up to $z=2$: \\ Sample properties and X-ray counterparts }

   \author{G. Toni\inst{1,}\inst{2}\thanks{\email{greta.toni4@unibo.it}}
   \and
          M. Maturi\inst{3,4}
          \and
          A. Finoguenov\inst{5}
          \and
          L. Moscardini\inst{1,}\inst{2,}\inst{6}
          \and
          G. Castignani\inst{1,}\inst{2}
          }

   \institute{Dipartimento di Fisica e Astronomia "A. Righi", Alma Mater Studiorum Universit\`a di Bologna, via Gobetti 93/2, 40129 Bologna, Italy
         \and
             INAF - Osservatorio di Astrofisica e Scienza dello Spazio di Bologna, via Gobetti 93/3, 40129 Bologna, Italy
             \and
             Zentrum für Astronomie, Universität Heidelberg, Philosophenweg 12, 69120 Heidelberg, Germany
             \and
             ITP, Universität Heidelberg, Philosophenweg 16, 69120 Heidelberg, Germany
\and
Department of Physics, University of Helsinki,
              Gustaf H\"allstr\"omin katu 2, 00560 Helsinki, Finland
              \and
             INFN - Sezione di Bologna, Viale Berti Pichat 6/2, 40127 Bologna, Italy
             }

   \date{Received -; accepted -}

% \abstract{}{}{}{}{} 
% 5 {} token are mandatory
 
  \abstract
   {}
   {We present a new galaxy cluster search in the COSMOS field through the use of the Adaptive Matched Identifier of Clustered Objects (AMICO). We aimed at producing a new cluster and group catalogue up to $z=2$, by performing an innovative application of AMICO with respect to previous successful applications to wide-field surveys, in terms of depth (down to $r<26.7$), small area covered ($\sim 1.69$ deg$^2$ of unmasked effective area) and redshift extent. This sample, and the comparative analysis we performed with the X-rays, allowed for the calibration of mass-proxy scaling relations up to $z=2$ and down to less than $10^{13} M_\odot$ and constitutes the base for the refinement of the cluster model for future applications of AMICO, like the analysis of upcoming Euclid data.
}
   {The AMICO algorithm is based on an optimal linear matched filter and detects clusters in photometric galaxy catalogues using galaxy location, photometric redshift and, in the simplest case, one galaxy property. We chose to use one magnitude as galaxy property, avoiding explicit use of galaxy colour for the selection of clusters. We used 3 different magnitudes by performing 3 independent runs in the $r$,$Y$ and $H$ bands using both COSMOS2020 and COSMOS2015 galaxy catalogues. We created a composite visibility mask, cluster models for the signal to detect and we estimated the noise directly from the data.}
   {We performed a matching of the catalogues resulting from the three runs and merged them to produce a final catalogue, which contains 1269 and 666 candidate clusters and groups with $S/N >3.0$ and $>3.5$, respectively. 490 candidates are detected in all three runs. Most of the detections unmatched within runs have $S/N <3.5$ which can be chosen as cut for a more robust sample. We assigned X-ray properties to our detections by matching the catalogue with a public X-ray selected group sample and by estimating, for unmatched detections, X-ray properties directly at the location of AMICO candidates based on Chandra+XMM-Newton data. There are in total 622 candidate clusters and groups with X-ray flux estimate. This large sample of candidates with X-ray properties allowed for the calibration of the scaling relations between two AMICO mass-proxies (richness and cluster amplitude) and X-ray mass and the study of their redshift dependence for the selection of the most stable photometric bands. 
   
   }
   {}

   \keywords{galaxies: clusters: general / galaxies: groups: general / galaxies: evolution / galaxies: luminosity function, mass function
               }
\titlerunning{AMICO-COSMOS galaxy cluster and group catalogue up to $z=2$}
\authorrunning{G. Toni et al.}
   \maketitle
%
%-------------------------------------------------------------------

\section{Introduction}
Clusters of galaxies are known as tracers of density peaks in the large-scale matter distribution and have proven to be powerful cosmological tools \citep[see][for a review]{allen11}. They can be used to constrain cosmological parameters through number density \citep[e.g.][]{rosati02,vikhlinin09,rozo10,costanzi19,lesci22a} and spacial distribution \citep[e.g.][]{veropalumbo14,marulli18,to21,lesci2022b}.
Additionally, galaxy clusters are important laboratories for the study of astrophysical processes, like those underlying galaxy formation and evolution. Clusters and groups as galactic environments influence the development of galaxy properties, as proven by the observed differences between field galaxies and galaxies in denser environments \citep[e.g.][]{dressler80,kuchner17,george11}. The understanding of galaxy cluster astrophysics is also fundamental to use them as cosmological probes. A reliable method for the identification of clusters and the determination of the \textit{mass-observable} relation linking the cluster masses to direct observables is a crucial requirement for the cosmological exploitation of galaxy clusters \citep[e.g.][]{pratt19,singh20}.

Galaxy clusters can be detected thanks to the hot gas galaxies are embedded in, which makes them bright X-ray sources \citep[e.g.][]{boehringer04,finoguenov10} and leaves an imprint at mm-wavelengths by distorting the Cosmic Microwave Background (CMB) spectrum (the thermal Sunyaev Zeldovich effect or SZ, \citeauthor{sz70} \citeyear{sz70}; e.g. \citeauthor{bleem15} \citeyear{bleem15}, \citeauthor{hilton18} \citeyear{hilton18}).
In the optical and near-infrared (NIR), clusters can be detected via the gravitational lensing effect on background sources \citep[e.g.][]{maturi05,stapelberg19,hamana20} and through the emission of member galaxies. Photometric catalogues of galaxies are largely used to detect clusters as overdensities of galaxies, by using different methods which exploit different galaxy properties. To mention some of the techniques widely used in literature: cluster red-sequence methods \citep[e.g.][]{rykoff14}, BCG methods \citep[e.g.][]{koester07}, wavelet filtering techniques \citep[e.g.][]{gonzalez14}, Voronoi tessellation methods \citep[e.g.][]{ramella01}, friends-of-friends algorithms \citep[e.g.][]{farrens11} and matched filters \citep[e.g.][]{postman96,bellagamba11}. 
Particularly challenging is the detection of galaxy clusters at high redshift which requires a proper choice of techniques and galaxy properties to be used. For instance, the dominance of red passive galaxies was shown to be less robust with increasing redshift, being $z \sim 1.4$ often found as a threshold for the presence of a fraction of star-forming galaxies more consistent with that of the field as well as more irregular galaxy morphologies \citep[e.g.][]{brodwin13,alberts16,strazzullo16}.

In this work, we performed a cluster search with the Adaptive Matched Identifier of Clustered Objects \citep[AMICO;][]{bellagamba18, maturi19}. This algorithm belongs to the class of linear optimal matched filters and it is capable of extracting signal with maximised signal-to-noise ratio ($S/N$).
AMICO has been chosen as one of the two algorithms for cluster detection officially adopted by the ESA \textit{Euclid} mission\footnote{\url{http://sci.esa.int/euclid/}} \citep{laureijs11}. The algorithm has distinguished itself in the context of the \textit{Euclid} Cluster Finder Challenge in terms of completeness and purity when applied to mock catalogues with the expected properties of \textit{Euclid} photometric catalogues \citep{euclid19}.
When compared to other methods, AMICO is characterised by the possibility to search for clusters with no need for spectroscopic information and without explicit use of colour as a galaxy property. This allows for cluster search in photometric catalogues up to high redshift reducing the possibility to bias the selection for presence (or absence) of cluster red-sequence. AMICO also includes an iterative detection and deblending procedure which allows for the detection of smaller and blended structures by removing the imprints of sequentially detected candidate clusters. The AMICO algorithm has already been successfully applied to wide-field surveys, like the Kilo-Degree Survey \citep[KiDS\footnote{\url{http://kids.strw.leidenuniv.nl/}};][]{dejong17} which gave origin to a cluster sample \citep{maturi19} already used for several cosmological \citep[e.g.][]{giocoli21,ingoglia22,lesci22a,lesci2022b} and cluster galaxy population studies \citep[e.g.][]{radovich20,puddu21,castignani22,castignani23}.

In this work, we searched for galaxy clusters with AMICO in the COSMOS 2-deg$^2$ field \citep{scoville07}.
The COSMOS survey offers the possibility to access high-quality multi-wavelength data up to high redshift. The availability of deep imaging and extremely accurate photometric redshifts gave us the chance to carry out this new and challenging application of the AMICO algorithm with respect to the ones successfully performed in the past. This cluster search with AMICO in the COSMOS field is innovative in terms, for instance, of redshift extent, depth and amount of area covered and allowed for the creation of a new catalogue up to $z=2$ and down to less than $10^{13} \, M_\odot$.

The catalogue of galaxy clusters and groups we produced is the result of three independent runs performed by using position, photometric redshift, and magnitude in one different band for each run. The multi-wavelength coverage of the COSMOS field 
allowed for a comparison with publicly available X-ray detected groups. In this work, we compared the AMICO-COSMOS catalogue with the catalogue presented in \citet{gozaliasl19} and exploited the availability of mass estimates to calibrate the scaling relations with AMICO mass proxies down to less than $ 10^{13} \, M_\odot$. We then repeated the same analysis by estimating X-ray mass with Chandra+XMM-Newton data at the locations of new detections as provided by AMICO and calibrated the scaling relations for a larger sample.

This paper is organised as follows.
In Section \ref{data} we introduce the galaxy catalogue we used as input for the cluster search. In Section \ref{amico} we present the fundamentals of the AMICO algorithm working principle. In Section \ref{methods} we describe the application of the AMICO algorithm to this peculiar survey configuration, going through the chosen cluster model, the creation of the composite visibility mask and of the noise. In Section \ref{results} we present the results of the cluster search, first by introducing the initial output of three AMICO runs, then by comparing them in order to create a final catalogue of candidates. Section \ref{xrays} is dedicated to the comparison with X-ray selected clusters and the calibration of a preliminary scaling relation based on successful matches. Section \ref{newdet} includes instead the analysis of the unmatched new detections for which we measured X-ray properties directly at the candidate locations and repeated the scaling-relation calibration with redshift dependence. In Section \ref{conclusions} we summarise the main results of this work. For the sake of simplicity, we favour the use of the term 'cluster' throughout the paper to refer to both candidates with the characteristics of galaxy clusters and of galaxy groups.
For this study, we assume a standard cosmology with matter density $\Omega_m=0.3$, dark energy density $\Omega_\Lambda =0.7$ and Hubble constant $h=0.7$.

%--------------------------------------------------------------------

\section{The COSMOS data sets}\label{data}

\begin{figure}
   \centering
   \includegraphics[width=8.8cm]{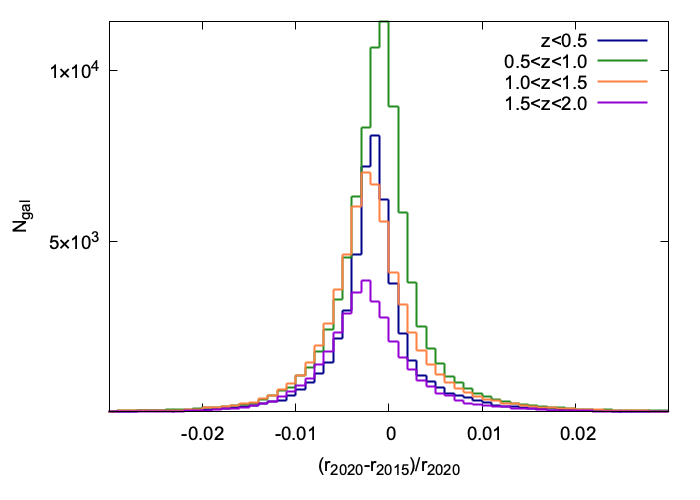}
 %     \caption{Distribution of galaxies present in both COSMOS2020 and COSMOS2015 catalogues as a function of the relative $r$ magnitude scatter ($\Delta r / r_{20}$). Different colours represent different redshift bins over the covered redshift range.}.
            \caption{Distribution of the relative magnitude scatter in the $r$-band for the galaxies
            present both in COSMOS2020 ($r_{2020}$) and in COSMOS2015 ($r_{2015}$) catalogues. Different colours represent different redshift bins, as labelled in the plot.}.
         \label{scatter1520}
\end{figure}

\begin{figure}
   \centering
   \includegraphics[width=8.8cm]{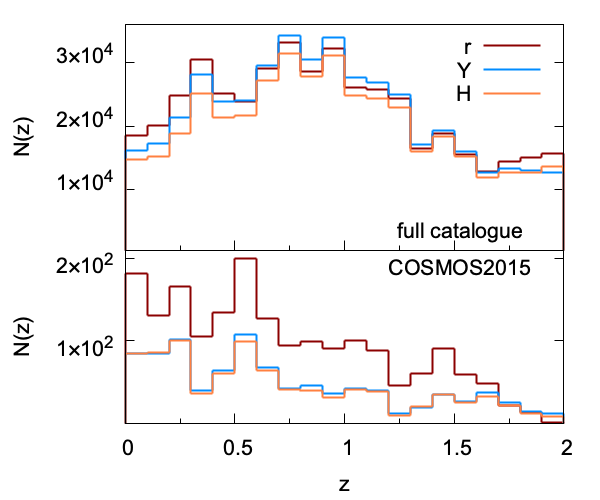}
      \caption{Distribution of selected galaxies as a function of redshift for the three runs in $r$, $Y$ and $H$-band magnitudes. \textit{Top panel}: full catalogues including selected galaxies of COSMOS2020 and the insert from COSMOS2015. \textit{Bottom panel}: only galaxies belonging to the insert from COSMOS2015.}
         \label{distr_z_peter}
\end{figure}

The Cosmic Evolution Survey (COSMOS) \citep{scoville07} has offered access over the years to a unique combination of deep data, with wavelength extension from radio to X-rays \citep[e.g.][]{koe07,zamojski07,cappelluti09,civano16,smolcic17} and large spectroscopic coverage \citep[e.g.][]{lilly07,hasinger18}. This has made this 2 deg$^2$-field source of large samples of galaxies characterised by photometric redshifts of extremely good quality up to high redshift. The field properties and accessibility made it ideal for studying the large-scale structure and the formation and evolution of structures in the Universe \citep[e.g.][]{hung16,cucciati18,laigle18}.

The photometric galaxy catalogue we used for our cluster search is the most recent release at the moment of writing: the COSMOS2020 catalogue \citep{weaver22}. We nevertheless integrated the catalogue with a sample coming from the previous release, COSMOS2015 \citep{laigle16}, due to missing photometry for some potential cluster galaxies. 
The COSMOS2015 catalogue comprises more than half a million galaxies extracted from the $zYJHK_s$ image, with data from Subaru \citep{taniguchi15} and VISTA \citep{mccracken12} telescopes via \texttt{SExtractor} dual-image mode \citep{bertin96}. Photometric redshifts, estimated using \texttt{LePhare} software \citep{arnouts02,ilbert06}, reach an accuracy better than $1\%$ at $z<1.2$ when compared with spectroscopic samples.

The main improvement in the COSMOS2020 catalogue with respect to the COSMOS2015 catalogue is the addition of new ultra-deep optical and NIR data from the Hyper Suprime-Cam (HSC) Subaru Strategic Program PDR2 \citep{aihara19} and from UltraVISTA survey Data Release 4 \citep{mccracken12,moneti23}. The use of new data is reflected in the nearly doubled number of detected sources and in the ability to reach the same photometric redshift uncertainty at nearly one magnitude deeper than in COSMOS2015. Moreover, the addition of the deep HSC-$i$ band to the detection image increases the completeness with respect to the previous release especially for small blue galaxies.

The COSMOS2020 release includes two different catalogues produced with independent extraction methods: (1) the \textsc{Classic} one which follows the same approach used in COSMOS2015, namely it is created with \texttt{SExtractor}, prior PSF-homogenisation; (2) \textsc{The Farmer} catalogue, produced by using purely parametric modelling with \texttt{The Tractor} \citep{lang16,weaver23}.  Both galaxy catalogues include in turn photometric redshifts computed with two different approaches, using \texttt{LePhare} and \texttt{EAZY} \citep{brammer08}.  \citet{weaver22} presented a detailed comparison between the different options and showed the very good consistency existing between the two estimation methods.

In this work, we chose to use the \textsc{Classic} catalogue in combination with photometric redshifts estimated with \texttt{LePhare}. This choice was motivated by the slightly larger effective area covered by this catalogue and by the better consistency with COSMOS2015. In this version of the catalogue, photometric redshifts were estimated using \texttt{LePhare} with a configuration similar to that used in \citet{ilbert13} and were found to reach percent accuracy at the brightest $i$ magnitudes up to a maximum value of $\sim 0.02(1+z)$ for $i<25$, with catastrophic failure of just a few percent.

The consistency in the source extraction and the redshift computation between COSMOS2020 and COSMOS2015 became crucial since we integrated the selected galaxy catalogue of the newest release with a sample of galaxies from COSMOS2015. We visually inspected different classes of objects flagged as masked and selected a group of galaxies, labelled with $\texttt{flag\_peter}=6$ or 4 in the catalogue by \citet{laigle16}, which were found to be bright and/or blended galaxies in dense environments, likely belonging to clusters. These galaxies are labelled with a general masking flag, not identified or have unsafe photometry in the newest release. 
Even if this galaxy sample from COSMOS2015 is statistically small, it contains typical cluster members, important for the cluster search.
Beside choosing the option with consistent photometric redshift estimation methods, we studied the consistency of the releases in terms of magnitude. We found a slight bias increasing with redshift which is nevertheless negligible within the analysed redshift range. For most of the galaxies present in both catalogue releases, the relative magnitude difference in the two galaxy samples is at sub-percent level, as shown in Figure \ref{scatter1520} for the $r$-band. Within the studied redshift range, this magnitude difference is negligible with respect to the resolution of the cluster model, as described in Section \ref{model}.

We performed three independent runs using three different magnitudes as galaxy properties.
Among the available bands, we chose to use HSC-$r$, UltraVISTA-$Y$ and UltraVISTA-$H$.
We considered only the UltraVISTA portion of the COSMOS field, namely, with RA[deg] $\in$ [149.30, 150.79] and Dec[deg] $\in$ [1.60, 2.81], where the bands used for source extraction are available as well as the bands chosen for this analysis. We searched for clusters over the unmasked fraction of this area, defined according to suited criteria (see Section \ref{mask}), for a final effective area of 1.69 deg$^2$.  We selected 3 final galaxy catalogues including all unmasked galaxies with available magnitude and photometric redshift. We discarded galaxies with anomalies in the redshift probability distribution, $p(z)$, namely with too wide or unrealistically peaked distributions which might bias the estimate of cluster properties. We performed a magnitude cut at the mode of each magnitude distribution defining the depth of the galaxy catalogues, as reported in Table \ref{galcat} . As a reference, the studied galaxy catalogue in the $r$-band extends to almost 3 magnitudes deeper than the catalogue used in the AMICO-KiDS cluster search \citep{maturi19}. The distribution in redshift of the selected galaxies in the three input catalogues is shown in Figure \ref{distr_z_peter}, where the top panel refers to the full input catalogues while the bottom panel refers to the galaxies added from COSMOS2015 only. Table \ref{galcat} summarises the properties of the used input galaxy catalogues, with the total number of galaxies from both releases (total number of unmasked galaxies in the input catalogue) and the number of galaxies among these which are coming from the COSMOS2015 sample.

   \begin{table}
      \caption[]{The selected input galaxy catalogues for the three analyses with the total number of galaxies belonging to both releases ($N_{gal,TOT}$) and the ones coming from COSMOS2015 ($N_{gal,2015}$).}
         \label{galcat}
     $$ 
         \begin{array}{c c c c c}
            \hline
            \noalign{\smallskip}
            Instrument-band      &  depth &  N_{gal,TOT} & N_{gal,2015}  \\
            \noalign{\smallskip}
            \hline
            \hline
            \noalign{\smallskip}
            HSC-r & < 26.7  & 450984 & 1854\\
            UltraVISTA-Y & < 26.1& 442449 & 934\\
            UltraVISTA-H &  <25.6 & 407067   & 883 \\
            \noalign{\smallskip}
            \hline
         \end{array}
     $$ 
   \end{table}
%

%-----------------------------

\section{The algorithm for cluster detection: AMICO}\label{amico}

The Adaptive Matched Identifier of Clustered Objects (AMICO) \citep{bellagamba18, maturi19} is a cluster detection algorithm based on a linear optimal matched filter \citep[e.g.][]{maturi05} which extracts a specific signal from a data-set affected by a noisy background, aiming at maximising the $S/N$. 
The data are modelled as the sum of a signal component and a noise component, accounting for cluster and field galaxies, respectively. Thus, the galaxy density $D(\boldsymbol{x})$, which is a function of the galaxy properties $\boldsymbol{x}$, can be written as $D(\boldsymbol{x}) = A M_c(\boldsymbol{x}) + N(\boldsymbol{x})$, where the signal component is expressed by an expected signal, namely the cluster model $M_c(\boldsymbol{x})$, scaled by the so-called \textit{amplitude} $A$, and $N(\boldsymbol{x})$ is the noise. The amplitude is computed as a convolution of the data with a kernel defined via a constrained minimisation which guarantees an unbiased and minimum-variance estimate of it.

The convolution filter $\Psi_c$, in case of  noise characterised by a white power spectrum, is expressed by $\Psi_c = \alpha^{-1}{M_c}/{N}$, that is the ratio between the cluster model and the noise.

Each galaxy in the catalogue is characterised by a sky position $\boldsymbol{\theta}_i$, a photometric redshift probability distribution $p_i(z)$ which in this case we model with a Gaussian distribution characterised by the mode and $1\sigma$ values, and an arbitrary number of additional galaxy properties. We focus here on the simple case of one single magnitude as galaxy property, so the set of considered properties for the $i$-th galaxy is $\boldsymbol{x_i}=( \boldsymbol{\theta}_i$, $p_i(z)$, $m_i)$, where $m_i$ is the galaxy magnitude. Given $\boldsymbol{\theta}_{i,c}$ the angular position of the $i$-th galaxy w.r.t. the cluster centre located in ($\boldsymbol{\theta}_c$, $z_c$), we can write the discretised form of the amplitude:

\begin{equation}
A(\boldsymbol{\theta_c}, z_c) = \alpha ^{-1}(z_c)\sum_{i=1}^{N_{gal}} \frac{M_c(\boldsymbol{\theta}_{i,c}, m_i) p_i (z_c)}{N(m_i,z_c)}  - B(z_c) \, ,
\end{equation}
where $\alpha$ is the amplitude normalisation and $B$ the average background contribution.
The expected variance of the amplitude, given by background stochastic fluctuations and Poissonian fluctuations generated by cluster members, is expressed by
\begin{equation}\label{variance}
    \sigma^2_A(\boldsymbol{\theta_c}, z_c) = \alpha^{-1}(z_c) + A(\boldsymbol{\theta_c},z_c)\frac{\gamma(z_c)}{\alpha^2(z_c)},
\end{equation}
where $\gamma$ is the cluster variance filter constant.
Once we express the typical redshift probability distribution for a galaxy located at $z_c$ as
\begin{equation}
q(z_c, z)=\frac{\sum_{i=1}^{N_{gal}}p_i(z-z_c+z_{p,i})p_i(z_c)}{\sum_{i=1}^{N_{gal}}p_i(z_c)} \, ,
\end{equation}
being $z_{p,i}$ the mode of the redshift distribution for the $i$-th galaxy,
the aforementioned filter constants representing amplitude normalisation, average background contribution and cluster variance, are respectively expressed by
\begin{equation}\label{fc1}
    \alpha (z_c) = \int \frac{M_c^2(\boldsymbol{\theta}-\boldsymbol{\theta_c},m,z_c)q^2(z_c,z)}{N(m, z_c)} \, d^2 \theta \, dm \, dz,
\end{equation}
\begin{equation}\label{fc2}
    B(z_c) = \alpha (z_c)^{-1}\int M_c(\boldsymbol{\theta}-\boldsymbol{\theta_c},\boldsymbol{m},z_c)q(z_c,z) \, d^2 \theta \, d m \, dz,
\end{equation}
\begin{equation}\label{fc3}
     \gamma(z_c)  =   \int \frac{M_c^3(\boldsymbol{\theta}-\boldsymbol{\theta_c},\boldsymbol{m},z_c)q^3(z_c,z)}{N^2(\boldsymbol{m}, z_c)} \, d^2 \theta \, d m \, dz.
\end{equation}

\subsection{Membership association}

Using a 3D grid with resolution $0.3^\prime$ on the sky plane and 0.01 in redshift, AMICO computes the map of amplitude and selects cluster candidates by looking for peaks in the map with the highest $S/N$.

AMICO determines a probabilistic membership for each galaxy once the cluster position is selected. The probability of the $i$-th galaxy of belonging to the $j$-th cluster is computed as

\begin{align}
    P_{i,j} = P_{F_i}\frac{A_j M_j (\boldsymbol{\theta}_i-\boldsymbol{\theta}_j, \boldsymbol{m}_i)p_i(z_j)}{A_j M_j (\boldsymbol{\theta}_i-\boldsymbol{\theta}_j, \boldsymbol{m}_i)p_i(z_j) + N(\boldsymbol{m}_i,z_j)} \, ,
\end{align}
where we account for possible previous associations to other clusters through the field probability, $P_{F_i}$. This latter has an initial value of 1 which decreases at each association and it is exploited as scaling factor to account for how much the galaxy is 'still available' for further assignments.

The membership probability is not only used to create a catalogue of cluster members, but it is also exploited to remove the imprint of detected clusters from the amplitude map with an iterative approach. This allows for the detection of blended and lower-$S/N$ candidates. The association probability is used to weight the contribution of members to the signal. 

\subsection{Mass proxies}
The membership probability assigned by AMICO to the cluster members is also used to compute two different cluster mass proxies in addition to the natural output of the filtering process, the amplitude, $A$. 
These proxies estimate the number of visible galaxies belonging to a cluster and are referred to as apparent and intrinsic richness. The former is simply the sum of all member probabilities of the $j$-th cluster:

\begin{equation}
    \lambda_j = \sum^{N_{gal}}_{i=1} P_{i,j} \, .
\end{equation}
This proxy is related to a direct observable but it is redshift dependent, since the number of visible galaxies decreases with distance.
The intrinsic richness follows the same approach but adding constraints on magnitude and distance from the cluster centre:
\begin{align}\label{lambdastardef}
    \lambda_{\star,j} = \sum^{N_{gal}}_{i=1}P_{i,j}
\,\,\,\, \text{with} \,\,\,\,
\begin{cases}
m_i < m_\star (z_j) + 1.5 \\
r_{i} < R_{200} (z_j)
\end{cases} \, ,
\end{align}
where $m_\star$ and the virial radius $R_{200}$ are parameters fixed by the chosen cluster model (see Section \ref{model}). The intrinsic richness, $\lambda_\star$, unlike the apparent one, has shown to be a nearly redshift independent mass proxy \citep{bellagamba18,maturi19}. 

\section{Applying AMICO to the COSMOS data}\label{methods}
We applied the AMICO algorithm to the COSMOS field by running three different and independent analyses, each one using one of the selected magnitudes. 
Despite AMICO can deal with an arbitrary number of quantities describing galaxy properties, we preferred to use one magnitude at a time, without therefore making explicit use of colour as galaxy property. This reduces the possibility to bias the selection for presence/absence of cluster red-sequence. We used the following bands: 
HSC-$r$, UltraVISTA-$Y$ and UltraVISTA-$H$, down to a magnitude of 26.7, 26.1 and 25.6, respectively, as reported in Table \ref{galcat}.
In all analyses, each galaxy of the input catalogue is considered as a data point with ID, sky coordinates, a single magnitude and an analytic photometric redshift probability distribution, $p(z)$, built with a Gaussian distribution characterised by $z_{peak}$ and  $z_{min}$, $z_{max}$, namely the $1\sigma$ errors.

\begin{figure}
   \centering
   \includegraphics[width=9cm]{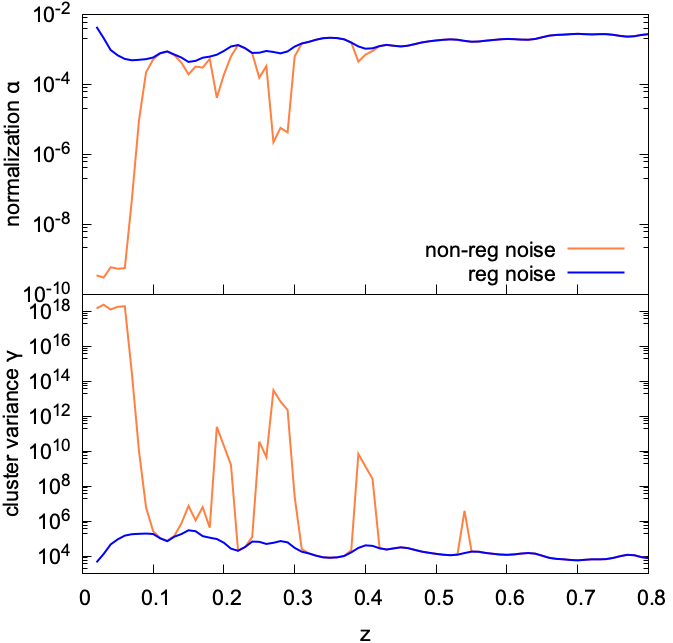}
      \caption{The effect of noise regularisation at $z<0.8$ on the redshift trend of two filter constants: the amplitude normalisation, $\alpha$ (top panel; see Eq. \ref{fc1}) and the cluster variance, $\gamma$ (bottom panel; see Eq. \ref{fc3}). The orange line represents a typical AMICO run on COSMOS data, using the $Y$-band magnitude, without noise regularisation while the blue line represents the exact same run with the noise assessment described in the text.}
         \label{noise_reg}
\end{figure}

\subsection{Cluster model}\label{model}
The cluster model describes the signal we expect to detect, namely the expected distribution of cluster galaxies in space and magnitude as a function of redshift. This can be constructed analytically from a radial profile $\Theta(r)$, where $r$ is the distance from the cluster centre, and from a luminosity function $\Phi(m)$:
\begin{equation}
M_c(r, m) = \Theta(r) \Phi(m) \, .
\end{equation}
For the radial distribution we used a NFW profile \citep{navarro97} with parameters from the scaling relation presented by \citet{hennig17}, where the properties of a sample of 74 SZ-selected massive clusters detected within the overlap between the Dark Energy Survey \citep[DES;][]{des05} and the South Pole Telescope survey \citep[SPT;][]{story13} were studied. The sample extends up to $z \sim 1.1$.
The estimate for the concentration parameter we adopted from this study is the mean full-population value $c=3.59$ and the typical virial mass of the model was chosen to be $M_{200}=10^{14} M_\odot$.

The luminosity function was instead assumed to follow a Schechter function \citep{schec76}
\begin{equation}
\Phi(m) \propto 10^{-0.4(m-m_\star)(\alpha+1)}\exp[{-10^{-0.4(m-m_\star)}]} \, .
\end{equation}
The faint-end slope, $\alpha$, was adopted as estimated by \citet{zenteno16} who analysed the 26 most massive clusters of the SPT survey sample and found a mean value for the full galaxy population of $\alpha=-1.06$. 
The characteristic magnitude, $m_\star$, and its redshift evolution was instead derived by evolving a typical massive elliptical galaxy with evolutionary synthesis models via the GALEV interface \citep{kotulla09}. For this model we relied on a Kroupa IMF and adopted a chemically consistent approach for a massive elliptical formed at $z=8$ with a high-redshift exponentially declining star-formation burst. We did this following the same approach as in \citet{castignani22,castignani23} and consistently with formation scenarios and passively evolving stellar populations in massive elliptical galaxies \citep[e.g.][]{holden04, delucia06, skelton12, castignani16}. The cluster model we built for this cluster search has a magnitude resolution of $\Delta m = 0.5$.\\

\subsection{Noise model}
The noise model, which accounts for field galaxies, can be approximated to the overall proprieties of the galaxy sample. This holds true under the assumption of negligible contribution of cluster galaxies.
Despite we observed overdensities localised in redshift in a noise model produced with the input galaxy catalogues, we found that the cause of these peaks is not to be attributed to physical overdensities (i.e. cluster galaxies) whereas it is likely due to photometric redshift accumulation points. The negligibility of cluster contribution to the noise for the COSMOS field was proven in two independent ways: (1) by removing galaxies attributed to groups with probability $>50\%$ up to $z=1$ according to the membership catalogue produced by \cite{george11}; (2) by taking the median of each noise pixel from non-overlapping portions of the field in order to attenuate the imprint of localised overdensities. Neither of these methods attenuated the observed peaks.

Given the small area covered by this cluster search and the small statistics of the galaxy sample with respect to past applications of AMICO to wide-field surveys, a noise regularisation was needed in order to get robust results.
To regularize the noise we attributed an arbitrary large value of noise to pixels without the contribution of any galaxy. This was necessary to make the integration space finite when computing the filter constants (see Eq. \ref{fc1}, \ref{fc2}, \ref{fc3}). Additionally, we assigned the same large value to pixels with magnitude $m< m_\star +3$. This was empirically found to successfully make the filter constants more robust.

Figure \ref{noise_reg} shows the influence that the performed regularisation has on two filter constants in comparison with a non-regularised noise model retrieved from COSMOS data sets. The regularisation of the noise successfully smooths out the largest fluctuations of the filter constants.

\begin{figure*}
   \centering
   \includegraphics[width=17cm]{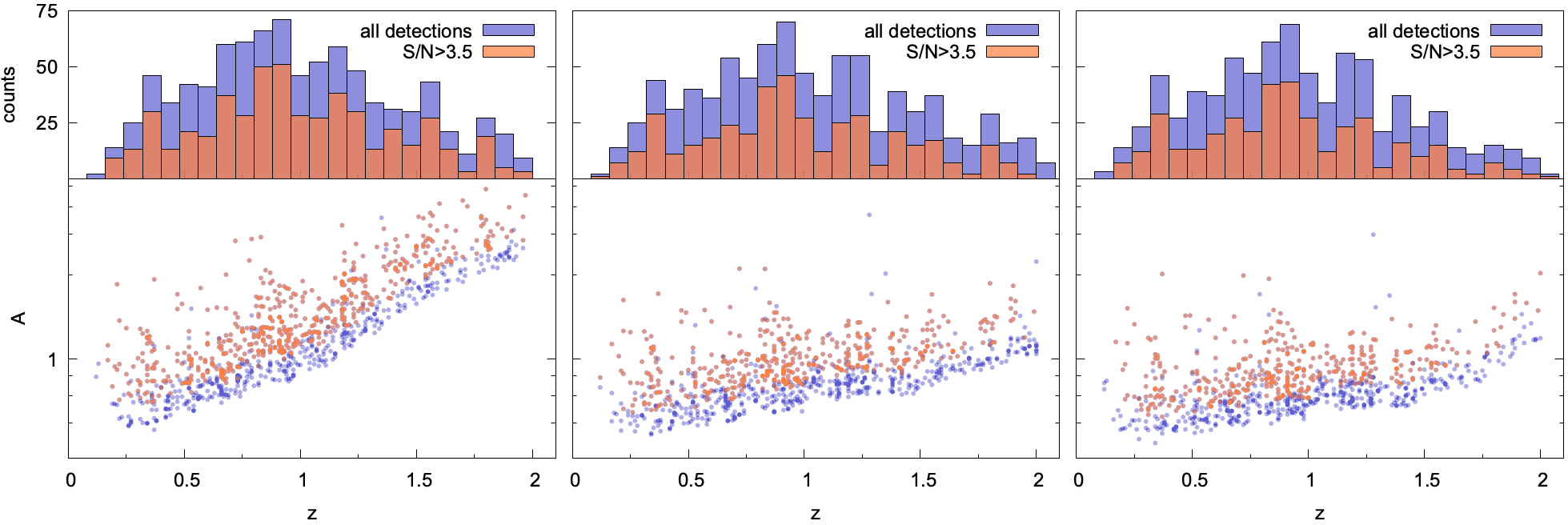}
      \caption{Number of clusters detected (top panels) and cluster amplitude (bottom panels) as a function of redshift for the initial cluster catalogues produced by analysis in $r$-band (left panels), $Y$- band (middle panels) and $H$-band magnitude (right panels). Orange histograms and points highlight detections with $S/N > 3.5$. }
         \label{statprop}
\end{figure*}

\subsection{Masked areas}\label{mask}

We created a visibility mask based on the input galaxy catalogue, as follows: (1) we masked the pixels where saturated stars fall, using the same sample \citet{weaver22} used for COSMOS2020. The sample is part of the Incremental Release of the HSC bright-star masks by \citet{coupon18}, extracted from the Gaia DR2 \citep{gaia18}, with magnitude threshold $G<18$. We masked different areas depending on the magnitude of stars in this range by simply dividing them in two groups, brighter or fainter than the median value; (2) we masked pixels lacking HSC unmasked objects in the full \textsc{Classic} catalogue in order to account for halos and diffraction spikes of bright foreground stars or other extended patterns; (3)
we masked pixels lacking galaxies according to the catalogue which includes inserts from COSMOS2015 to account for these additional galaxies coming from the previous release.
Masked pixels combining these criteria were used to create the final composite visibility mask. The final effective area on which the cluster search was performed is of 1.69 deg$^2$.

%------------------------------------------------------------------------------------------------------RESULTS

\section{The cluster and group catalogues}\label{results}

In this Section we present the resulting catalogues of the cluster search performed in COSMOS with the AMICO algorithm. The candidate samples are then matched and compared and a few examples of detections are discussed.

\subsection{Results of the cluster search}

We performed a cluster search over the effective area of 1.69 deg$^2$ in the COSMOS-UltraVISTA field as previously described. We chose $(S/N)_{min}=3.0$ and cut the catalogues at $\lambda_\star >1$. We produced three initial cluster catalogues from the three analyses performed by using different photometries: \\

\noindent\textbf{\textit{r} - band}: we detected 893 clusters in the range $0.1<z<2.0$ by using the HSC-$r$ magnitude as galaxy property. Among these, 514 candidates ($\sim 58\%$) were detected with $S/N >3.5$; \\
\noindent\textbf{\textit{Y} - band}: we detected 845 clusters in the range $0.1<z<2.0$ by using the UltraVISTA-$Y$ magnitude. Among these, 408 candidates ($\sim 48\%$) were detected with $S/N >3.5$; \\
\noindent\textbf{\textit{H} - band}: we detected 786 clusters in the range $0.1<z<2.0$ by using the UltraVISTA-$H$ magnitude. Among these, 382 candidates ($\sim 49\%$) were detected with $S/N >3.5$. \\

\noindent Figure \ref{statprop} shows the distribution of the number of detections and their amplitude as a function of redshift. The minimum detectable amplitude increases as a function of redshift since the further we observe, the harder it is to detect small clusters.
Differences between the number of detected clusters and cluster properties between different runs can be due to the differences in depth and availability of galaxies in the input catalogues. This is also visible in the general trend of amplitude with redshift, which especially at $z>1$ increases faster and up to larger values for the $r$-band catalogues with respect to the other catalogues, due to the larger number of available galaxies in the input catalogue.

The resulting cluster catalogues include: identification number, sky coordinates, redshift, $S/N$, amplitude $A$, apparent richness $\lambda$, intrinsic richness $\lambda_\star$, likelihood, fraction of the cluster which is masked, cluster redshift probability distribution and redshift uncertainty. Beside the cluster catalogue, AMICO creates the list of galaxy members for each detected cluster with their membership and field probability.

\begin{figure*}
   \centering
   \includegraphics[width=18cm]{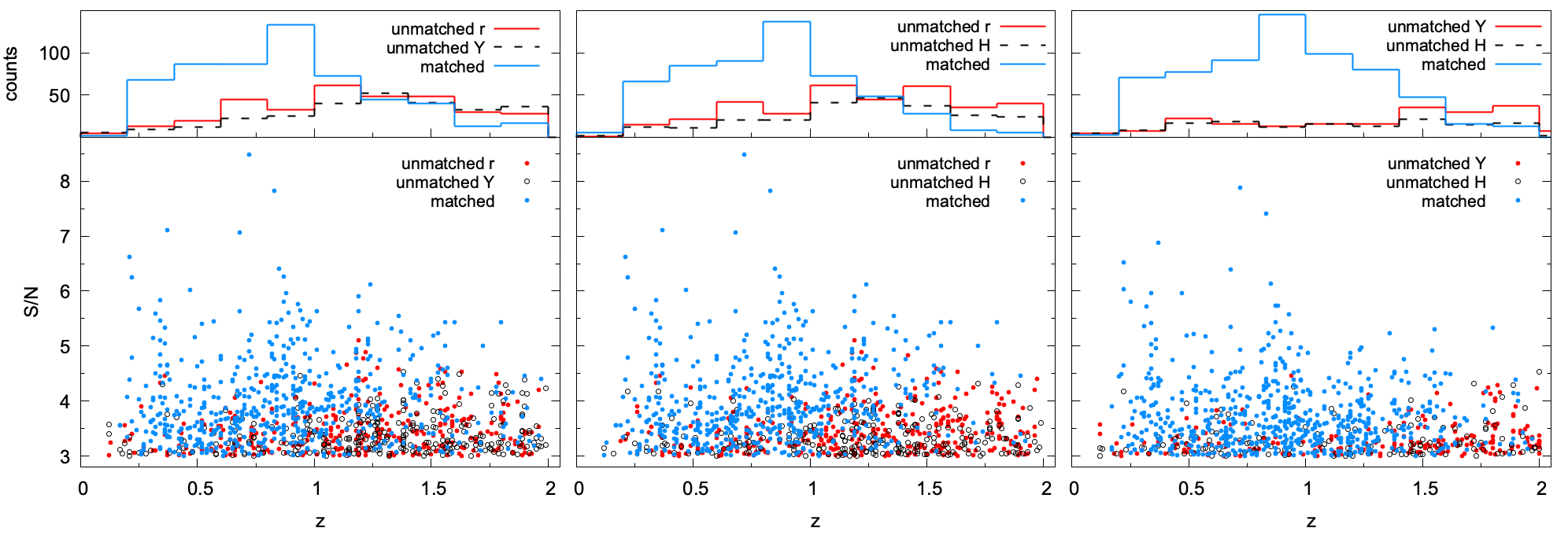}
      \caption{Number of detected clusters (top panels) and their $S/N$ (bottom panels) as a function of redshift for matched (blue) and unmatched (red) detections within AMICO runs. From left to right: matching $r$ vs $Y$, $r$ vs $H$ and $Y$ vs $H$. Filled red dots and solid red lines indicate the first catalogue unmatched detections, while empty black dots and dashed lines, the second one.}
         \label{snr}
   \end{figure*}

\subsection{Matching the r,Y,H catalogues}\label{rYH}
We performed a pairwise three-dimensional matching between the three catalogues adopting d$z$=0.05(1+$z$) as redshift separation and d$rad$ = 0.5 $Mpc/h$ as physical sky separation. These values were chosen according to the uncertainties of cluster redshift and position estimates and were tested by visual inspection of preliminary matching results. In the matching procedure we used an a-priori sorting of the input catalogues by amplitude, $A$.

\begin{figure}
   \centering
   \includegraphics[width=8cm]{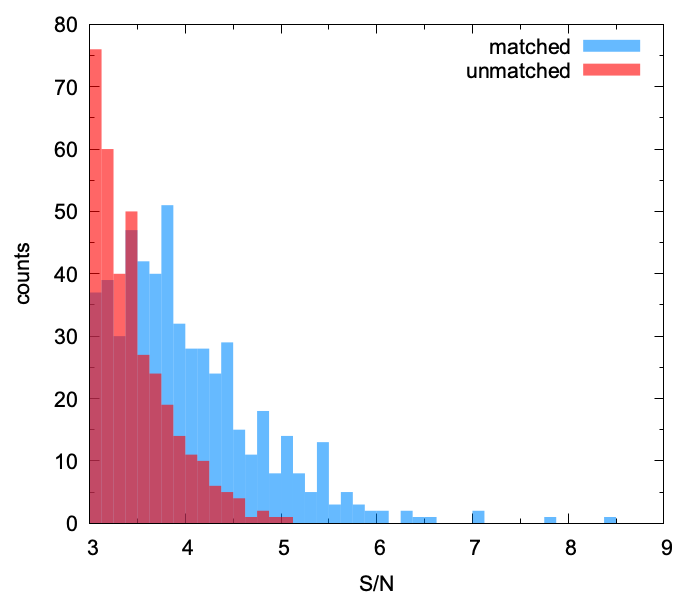}
      \caption{Distribution of matched (blue) and unmatched (red) cluster detections in the comparison $r$ vs $H$, as a function of signal-to-noise ratio. Most of the unmatched detections have $S/N <3.5$.}
         \label{snrdistr}
   \end{figure}

We found correspondence between 561 candidate clusters detected in the $r$ and $Y$ analyses, namely around 63\% and 66\% of the two catalogues, respectively. 
Within the $Y$ and $H$ analyses, 642 detections found a match, namely around 76\% and 82\% of the two catalogues, respectively.
542 candidate clusters were matched when comparing $r$ and $H$ analyses, namely around 61\% and 69\% of the two catalogues, respectively.
The results of this matching analysis are displayed in Figure \ref{snr}, where we show the redshift distribution and the $S/N$-$z$ of matched and non-matched detections for the three different pairs of catalogue combinations.
As visible in Figure \ref{snr} and as highlighted by the $S/N$ distribution of matches and non-matches for the $r$ vs $H$ comparison in Figure \ref{snrdistr} (which is the comparison with the highest number of non-matches), the majority of detections not having correspondence between runs have $S/N<3.5$, showing how this can be adopted as a reasonable and conservative $S/N$-cut for the catalogue in order to have a more robust sample.

With a two-way matching between different combinations of the three catalogues and by adopting the same procedure described before, we found a total number of 490 candidate clusters and groups detected independently in all the three analyses. 
Given the results of the matching between catalogues resulting from the three runs, we created a final catalogue containing all candidates, both matched and unmatched (we counted as one the detections successfully found in more than one run according to our matching). Our final catalogue contains a total number of 1269 candidate clusters and groups, given that all runs were performed with $(S/N)_{min}=3.0$. Among these, 666 candidates have been detected with $S/N>3.5$ in at least one of the runs. 

When referring to cluster properties of specific detections hereinafter we report the mean of the values obtained in the three different runs, unless otherwise specified.

\begin{figure*}
   \centering
   \includegraphics[width=16cm]{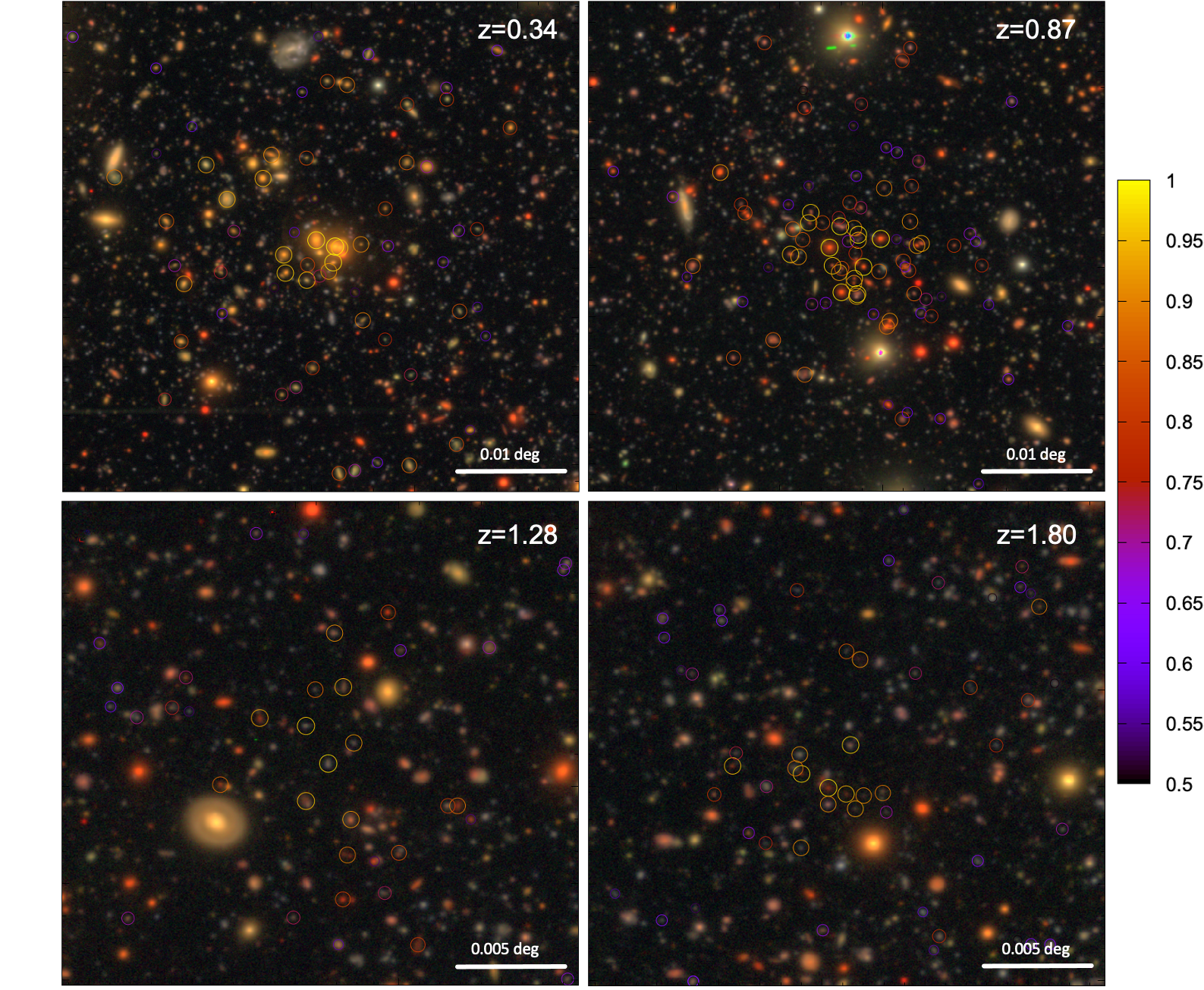}
      \caption{Four examples of detections at different redshifts as identified in the $r$-band run. Stamps are HSC $g,r,i$ colour-composite images centred in the centre of the cluster as identified by AMICO, with a side of 0.05 $deg$ for the top panels and 0.025 $deg$ for the bottom panels. Circles mark the associated galaxies, sized and coloured according to their association probability (as in the bar on the right).  \textit{Top left}: candidate at $z=0.34$ with $\lambda_\star \sim 17$; \textit{top right}: candidate at $z=0.87$ with $\lambda_\star \sim 27$; \textit{bottom left}: candidate at $z=1.28$ with $\lambda_\star \sim 26$; \textit{bottom right}: candidate at $z=1.80$ with $\lambda_\star \sim 40$.}
         \label{example1}
\end{figure*}

\begin{figure}
   \centering
   \includegraphics[width=8.5cm]{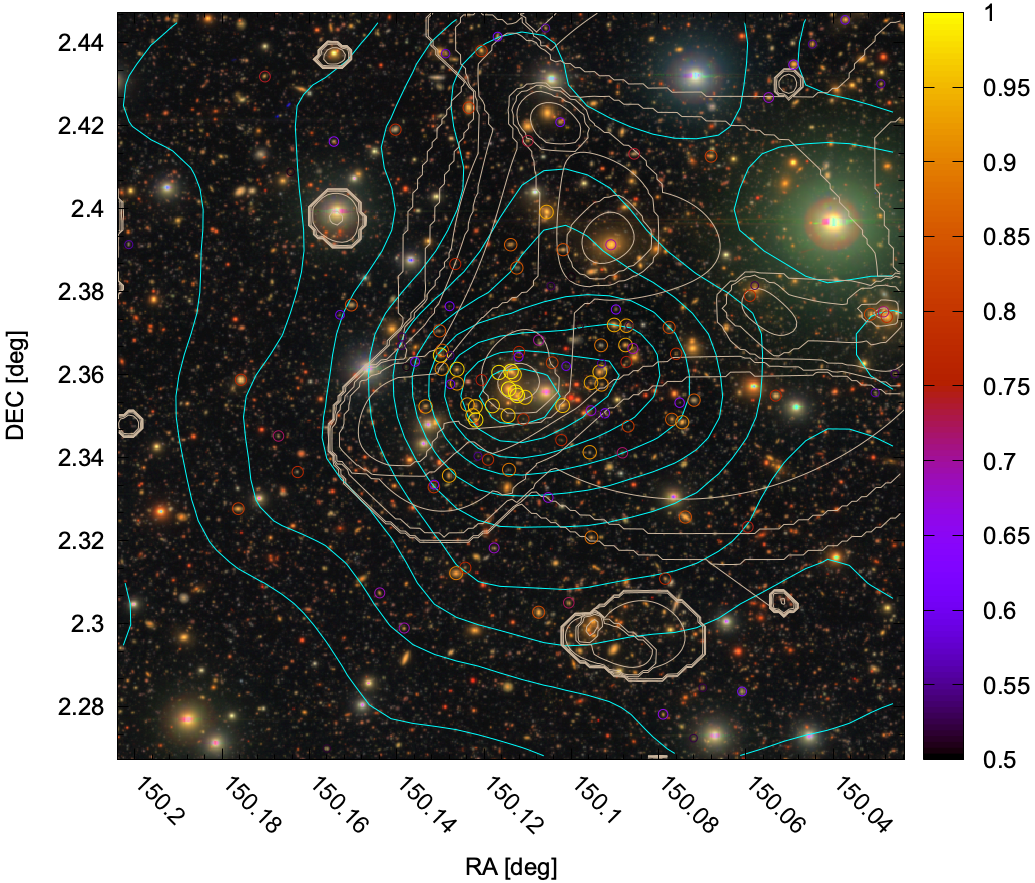}
      \caption{The assembly of cluster substructures at $z=0.22$ found by \citet{smolcic07} via radio-BCG, as detected by AMICO. Circles indicate assigned members with colour-coded probability (as in the bar on the right) and are overlaid with AMICO amplitude and X-ray contours (cyan and white, respectively).}
         \label{supercluster}
\end{figure}

\begin{figure}
   \centering
   \includegraphics[width=8.5cm]{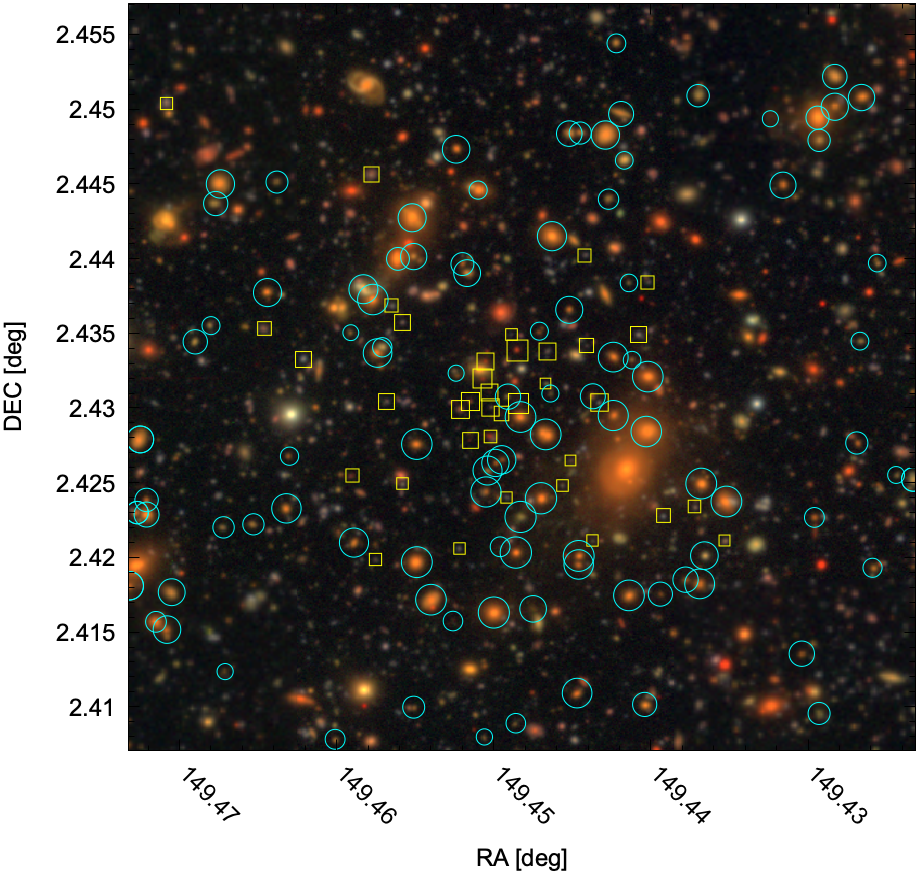}
      \caption{Couple of detections aligned along the line-of-sight, as detected in the $Y$-band run. Members of the foreground cluster, located at $z = 0.47$, are indicated by cyan circles. Yellow squares indicate instead background cluster galaxies ($z=1.56$). Size of circles/squares is proportional to the membership probability. The HSC $g,r,i$ composite image has a size of 0.05 deg and is centred in the centre of the background cluster.}
         \label{overlappingclusters}
\end{figure}

\subsection{Examples of detections}
We visually inspected the location of a sample of detected clusters in the optical colour-composite images from HSC DR3 \citep{aihara22}. Figure \ref{example1} shows 4 randomly picked detections as detetcted in the $r$-band, at different redshifts, with $S/N>4.0$.
Among the candidate clusters, we successfully detected the main cluster of the system identified by \citet{smolcic07} via wide-angle-tail radio galaxy. \citet{smolcic07} found the radio galaxy to coincide with an elliptical galaxy which was identified as the brightest cluster galaxy (BCG) of a cluster at $z \sim 0.22$. We detected this cluster in all bands and we identified 146 galaxy members (in the $Y$-band) associated with probability $>50 \%$, including the BCG (radio galaxy) which was assigned with the highest probability and has a position consistent with the centre of the cluster as found by AMICO. The structure, as detected by AMICO, is shown in Figure \ref{supercluster}, with associated members, amplitude and X-ray Chandra+XMM-Newton (\citeauthor{gozaliasl19} \citeyear{gozaliasl19}; see Section \ref{xrays}) contours.
We were not able to distinguish the assembly of more than one cluster as stated by \citet{smolcic07}, probably because of several saturated stars affecting the concerned region. 
Nevertheless, among the brightest galaxies associated to the cluster, we found at least 3 galaxies located in correspondence of the substructures, coinciding with the peaks of the diffuse X-ray emission. This structure has been identified by AMICO as a single cluster but the presence of these bright galaxies far from the cluster centre can indicate the assembly of more substructures.

We compared our list of candidates with other group and cluster catalogues available for the COSMOS field. Since the catalogues have been retrieved with different methods and therefore present different definitions of richness and quality of the detections, it is not straightforward to make consistent comparisons and considerations. We simply include in our final catalogue the corresponding identification numbers for the 10, 182, 307, 7, 11 and 585 detections matched within d$z$=0.05(1+$z$) and d$rad$ = 0.5 $Mpc/h$, from the catalogues by \citet{zatloukal07}, \citet{knobel09}, \citet{knobel12}, \citet{castignani14}, \citet{iovino16} and \citet{sarron21}, respectively.

One of the goals of this study was to search for high-redshift clusters.
We detected 273 clusters at $z>1.5$, among which 125 were detected with $S/N >3.5$ in at least one run.
We found that 111 common identifications with the catalogue by \citet{sarron21} have $z>1.5$, 3 of which are also detected by \citet{zatloukal07}.

\paragraph{\textit{Clusters overlapping along the line-of-sight.}}
We performed a two-dimensional matching on our final catalogue (without using redshift information) and we found 269 sets\footnote{pairs or groups of more than two detections} of detections lying within 0.02 deg from each other. Some of these cases are in a configuration that makes it challenging to detect them, therefore the background objects are often not known in literature. For instance, among these we found that 116 sets of detections (of which 96 are pairs) have a sky separation $\leq 0.01$ deg, involving a total number of 253 objects that are almost in perfect alignment along the line-of-sight with another object (or more than one). An example of this kind of detection pair is shown in Fig. \ref{overlappingclusters}, where an already known cluster at $z= 0.47$ is virtually aligned with a background cluster at $z = 1.56$, not included, for instance, in most of optical and X-ray cluster catalogues.
The identification of these sets of detections is important in this cluster search, where the number of detected objects per square degree is high. In particular, having several detections close to each other on the sky plane can bias the estimation of X-ray properties. This is why we used the information about the position of these sets of detections to clean the catalogue for the calibration of scaling relation described below, in Section \ref{xrays}.

\subsection{Spectroscopic counterparts}
We assigned spectroscopic counterparts to cluster members associated by AMICO by making use of a sample of galaxies from 13 public spectroscopic surveys (zCOSMOS-b, \citeauthor{lilly07} \citeyear{lilly07}; PRIMUS, \citeauthor{coil11} \citeyear{coil11}, \citeauthor{cool13} \citeyear{cool13}; GEEC2, \citeauthor{balogh14} \citeyear{balogh14}; FORS2, \citeauthor{comparat15} \citeyear{comparat15}; DEIMOS, \citeauthor{hasinger18} \citeyear{hasinger18}; VIS3COS, \citeauthor{paulinoafonso18} \citeyear{paulinoafonso18}; hCOSMOS, \citeauthor{damjanov18} \citeyear{damjanov18}; FMOS-COSMOS, \citeauthor{kashino19} \citeyear{kashino19}; C3R2, \citeauthor{masters19} \citeyear{masters19}; MUSE, \citeauthor{rosani20} \citeyear{rosani20}; LEGA-C, \citeauthor{vanderwel21} \citeyear{vanderwel21}; MAGIC, \citeauthor{epinat21} \citeyear{epinat21}, \citeauthor{abrilmelgarejo21} \citeyear{abrilmelgarejo21}, Epinat et al., in prep.; DESI, \citeauthor{adame23} \citeyear{adame23}) collected in the COSMOS Spectroscopic Redshift Compilation (Khostovan et al., in prep.). We found that 612, 741 and 720 detections have at least one member associated with association probability $>50\%$ that has a spectroscopic redshift, in the $r$, $Y$ and $H$-band analyses, respectively. Among these, 373, 449 and 454 (i.e. 61\%, 61\% and 63\%) have more than 3 members with spectroscopic redshift, respectively. All these cluster candidates have their cluster redshift assigned by AMICO compatible with $z_{spec}$, which is defined as the mean of spectroscopic redshift of associated members, being $\Delta z /(1+z_{spec})<0.03$. If we consider the final catalogue obtained by matching detections between the three different runs (as described in Sect. \ref{rYH}), the total effective number of detections with mean spectroscopic redshift based on more than 3 member galaxies is 567 and all of them have mean spectroscopic redshift compatible with that assigned by AMICO.

%----------------------------------
%% XRAY comparison

\section{Comparison with X-ray group catalogue}\label{xrays}
We based our comparative analysis of the AMICO detections with X-ray selected groups on the work by \cite{gozaliasl19}. In this work a catalogue of galaxy groups was produced for the COSMOS field by using Chandra \citep{elvis09,civano16} and XMM-Newton \citep{hasinger07,cappelluti09} $0.5-2.0$ keV data combined via wavelet transform \citep{vikhlinin98}. 

The catalogue contains 247 groups covering the range $0.08\leq z<1.53$ and with masses of $M_{200}=8 \times 10^{12} \, - \, 3 \times 10^{14} \, M_\odot$. Virial masses were estimated via the scaling relation derived through stacked weak lensing analysis by \cite{leauthaud10}. This X-ray selected catalogue is a revised and extended version of the COSMOS group catalogues presented in \citet{finoguenov07} and \citet{george11}. To optically validate the X-ray extended sources and estimate their redshifts, both spectroscopic \citep{hasinger18} and photometric \citep{ilbert09,laigle16,mccracken12} galaxy samples were used. Data reduction and member identification, performed via \textit{refined red-sequence} approach, are described in details in \citet{george11} and \citet{gozaliasl19}.

We performed a three-dimensional matching within d$z$=0.05(1+$z$) and d$rad$ = 0.5 $Mpc/h$. We found: 104 successful matches for the $r$-band run, 107 matches for the $Y$-band run and 99 matches for $H$-band one. If we consider AMICO detections matched within different runs, the total number of detections with correspondence in the X-ray catalogue by \citet{gozaliasl19} is 122, that is the 55\% of the X-ray detections. For the comparison, we considered consistent volumes occupied by the cluster searches, namely accounting for different redshift extent and effective unmasked areas covered. This comparison displayed a general good matching quality with most of the successfully paired detections lying within 0.1 $Mpc/h$ and 0.02 of redshift scatter. 
We found a slight and negligible redshift bias, namely redshift estimates for AMICO are slightly smaller with respect to X-ray identifications, with an average bias of $\Delta z/(1+z) \approx -0.002 \pm 0.001$.

\subsection{Quality flags}
\cite{gozaliasl19} included 4 quality flags in their cluster catalogue. Flag 1 labels the safest detections, with spectroscopic members; flag 2 is for cases of fore-/background contamination; flag 3 marks lack of spectroscopic counterparts; flag 4 is assigned to the least safe sample, with ambiguous optical counterpart association. In our comparative analysis, we found that 77\% of the successful matches with AMICO catalogues are part of the safest X-ray sample. The safest sample of X-ray detections has more successful than unsuccessful matches while for the clusters flagged with 2, 3 and 4 the opposite is true: around the 57\% of problematic detections in the X-rays (flag 2, 3 or 4) do not pair with any AMICO detection. This can be read as an indication of the reliability of the flagging system by \cite{gozaliasl19} and therefore the general reliability of the safest X-ray candidate clusters (flag 1).

\subsection{The $r,Y,H$ analyses when compared with X-rays}
A total number of 86 clusters have been found in all three AMICO catalogues and successfully matched with the catalogue by \cite{gozaliasl19}. We can take the X-ray catalogue as a reference to make considerations on the different runs. We need to take into account only the common volume, so we restrict this comparison to the redshift range covered by the X-ray catalogue. The catalogue produced by the $Y$-band run and $r$-band run appears to be more complete with respect to the X-ray reference, having 107 and 104 matched clusters, respectively. When we consider instead the relative number of matches in the samples of 'new' detections, that are found by AMICO only in that specific run, we see that the $r$-run matches nevertheless only $\sim 4\%$ of its new detections, whereas $\sim 7\%$ of new $Y$ and $H$ detections are paired in the X-rays.

\begin{table}
      \caption[]{Mass-proxy scaling relation parameters and pivot values (see Eq. \ref{bg_sr}) based on detections matched with the group catalogue by \citet{gozaliasl19}.}
         \label{scalrel}
     $$ 
         \begin{tabular}{c c c c}
            \noalign{\smallskip}
            \textbf{\textit{r}-band} & & &  \\
            \hline
            \hline
             &   $\alpha$ & $\beta$ & $O_{piv}$ \\
            \noalign{\smallskip}
            \hline
            \noalign{\smallskip}
            $\lambda_\star$ & $-0.320 \pm 0.025$ & $0.780 \pm 0.129$ & 21\\
            \noalign{\smallskip}
             $A$  & $-0.311 \pm 0.025$ & $0.917 \pm 0.151$ & 1.4\\
            \noalign{\smallskip}
            \hline
            \hline
            & & & \\
            \noalign{\smallskip}
            \textbf{\textit{Y}-band} & & &  \\
            \hline
            \hline
             &   $\alpha$ & $\beta$ & $O_{piv}$ \\
            \noalign{\smallskip}
            \hline
            \noalign{\smallskip}
            $\lambda_\star$ & $-0.351 \pm 0.026$ & $0.961 \pm 0.146$ & 15\\
            \noalign{\smallskip}
             $A$  & $-0.330 \pm 0.026$ & $1.155 \pm 0.198$ & 1.0\\
            \noalign{\smallskip}
            \hline
            \hline
            & & & \\
            \noalign{\smallskip}
            \textbf{\textit{H}-band} & & &  \\
            \hline
            \hline
             &   $\alpha$ & $\beta$ & $O_{piv}$ \\
            \noalign{\smallskip}
            \hline
            \noalign{\smallskip}
            $\lambda_\star$ & $-0.352 \pm 0.027$ & $1.027 \pm 0.145$ & 15\\
            \noalign{\smallskip}
             $A$  & $-0.341 \pm 0.027$ & $1.132 \pm 0.180$ & 0.9\\
            \noalign{\smallskip}
            \hline
            \hline
         \end{tabular}
     $$ 
   \end{table}

\subsection{Mass-observable scaling relations}\label{scalingfit}

We studied the relation between mass proxies provided by AMICO and X-ray derived mass for the sample of successfully matched clusters.

\begin{figure*}
   \centering
   \includegraphics[width=18cm]{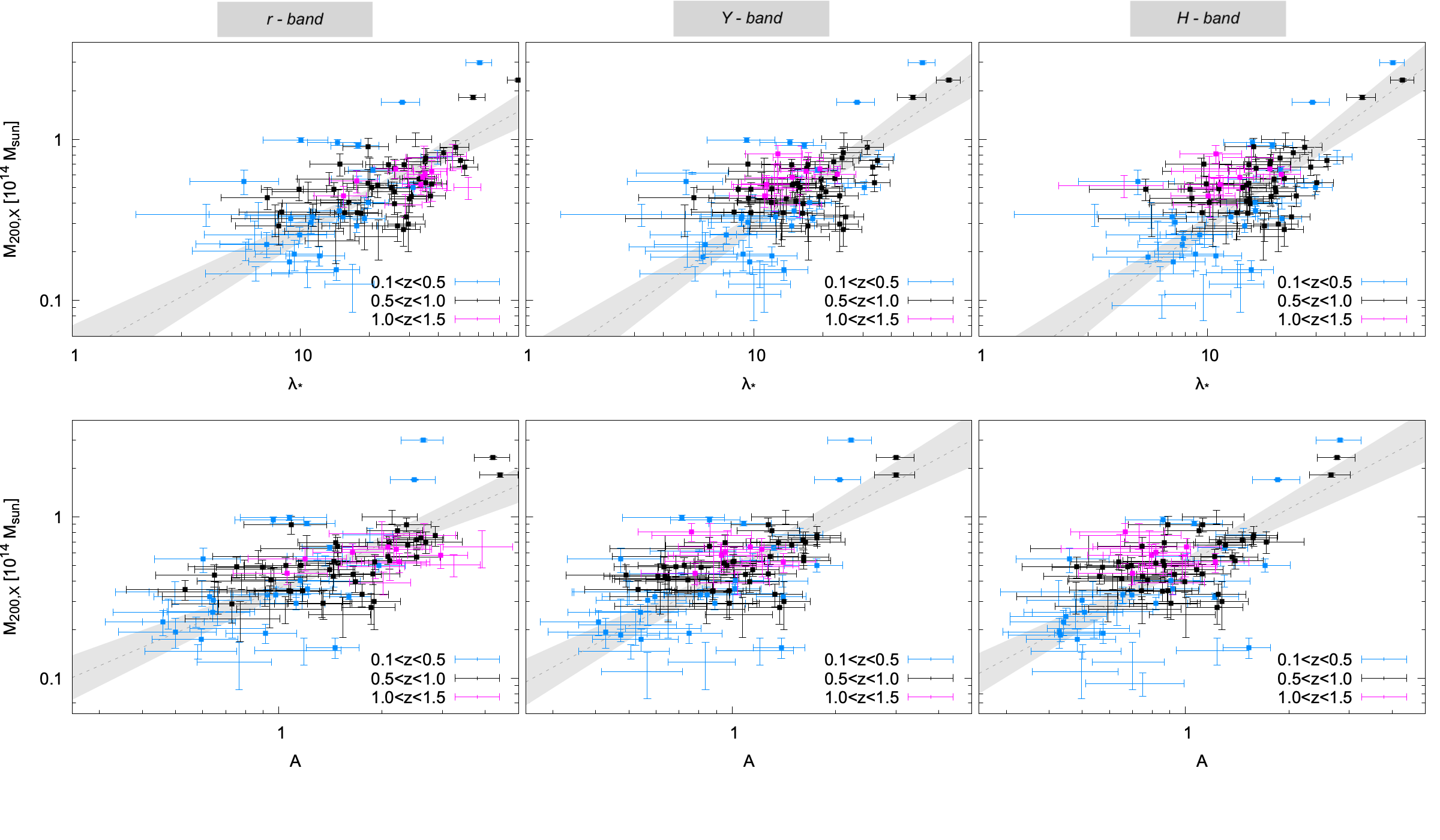}
      \caption{Relation between X-ray mass and richness (top panels) or amplitude (bottom panels) for the X-ray matched clusters in the catalogue by \citet{gozaliasl19} in the range $z \in $ [0.1, 1.5]. From left to right: detections retrieved from analysis in $r$, $Y$ and $H$-band. Different colours refer to different redshift bins, as labelled in the plot. Grey dashed line and shaded area indicate the best-fitting relation with the corresponding 1$\sigma$ errors. Data points marked by squares are the detections belonging to the safest sample in the X-ray catalogue and without fore-/background or nearby contamination. Error-bar points without central square are matched detections not used for the scaling relation calibration.}
         \label{scalrelfig}
\end{figure*}

The X-ray catalogue by \cite{gozaliasl19} comprises the 0.1-2.4 keV rest frame X-ray luminosity ($L_X$) measured within $R_{500}$ and the virial mass $M_{200}$ estimated through the scaling relation presented by \cite{leauthaud10}.
We studied the relation between X-ray virial mass (and luminosity) with both AMICO amplitude ($A$) and intrinsic richness ($\lambda_\star$). 
For this analysis we used only detections flagged as safe and with spectroscopic counterparts (flag 1 in \citet{gozaliasl19}) which do not fall within $0.02$ deg from another detection with $\lambda_\star > 20$ and identified with $S/N>3.5$. This reduces the possibility of introducing outliers due to projection effects and contamination of a nearby object.

Mass-observable calibration from AMICO mass-proxies was performed for the AMICO-KiDS cluster sample \citep{maturi19} via stacked weak-lensing analysis by \citet{bellagamba19}. We used the scaling relation expression used by \cite{bellagamba19} to fit the data for the matched cluster sample:
\begin{equation}\label{bg_sr}
    \log \frac{M_{200}}{10^{14}M_\odot} = \alpha + \beta \log\frac{O}{O_{piv}}  + \gamma \log\frac{E(z)}{E(z_{ref})} \,  \, ,
\end{equation}
being $O$ the AMICO mass proxy and $O_{piv}$ its median value in the considered sample. In this part of the analysis, we neglected the redshift dependency term of the relation ($\gamma =0$), since the limited size of our sample does not allow for robust analyses of redshift evolution.
The fitting analyses were performed by taking into account the $1 \sigma$ errors for X-ray mass, the amplitude variance (Eq. \ref{variance}) and the square root of $\lambda_\star$. We made use of the R package for LInear Regression in Astronomy (\texttt{LIRA}) described in \citet{sereno16}. The best-fit parameters are reported in Table \ref{scalrel} and the corresponding relations shown in Figure \ref{scalrelfig}. 
A defined relation is visible in almost all combinations among the catalogues, with a generally smaller scatter for the results of the $r$-band run with respect to the analyses using other photometries. The large scatter is expected for a sample extending over this wide range of masses. Detections marked with squared points in Fig. \ref{scalrelfig} are the detections used for the scaling relation fit, namely belonging to the safest sample detected in the X-rays, with reduced chances to have fore-/background contamination and they all have spectroscopic members. The studied sample extends down to low masses, with the least massive matched cluster having $M_{200} \approx 9.3 \times 10^{12} M_\odot$ and $L_{X} \approx 3.3 \times 10^{41} erg/s$.

\section{X-ray counterparts for new detections}\label{newdet}
After matching AMICO detections and X-ray selected groups by \citet{gozaliasl19}, we studied the sample of non-matched detections and look for their possible X-ray counterparts. We measured X-ray properties for these detections directly at their locations as identified by AMICO.

We used the combined Chandra and XMM-Newton maps of emission residuals in the 0.5--2 keV band after removing instrumental and sky background as well as unresolved X-ray emission. XMM-Newton provides approximately 70\% of the sensitivity in the combined data. For testing the effect of the total flux including the unresolved emission, we only used the XMM-Newton data, as it is the one that suffers from the resolution effects. The data reduction in producing the maps is identical to the analysis published by \citet{gozaliasl19}.
Using the redshifts of the AMICO clusters, we estimated the count rates inside the 200 kpc radius and in obtaining the X-ray properties we extrapolated those numbers to the iteratively estimated $R_{500}$ radius, following the procedure outlined in \citet{finoguenov07}. The size of the aperture matches well the expected size of the emission zones of the AMICO clusters found below the limit of the published X-ray catalogs in COSMOS, and so the flux extrapolations are minimal, while the overlap in the extraction zones between neighbouring clusters is not significant at $z>0.2$. 

We kept only detections with significant X-ray emission by cutting at flux significance above the 1$\sigma$ limit. The total number of new detections with X-ray flux estimate above the significance limit is 500 out of the 1147 analysed candidates ($\sim 44\%$). If we consider only objects detected by AMICO with $S/N>3.5$, 267 have significant X-ray emission out of the total 577 analysed candidates ($\sim 46\%$).

By adding together successful matches with the X-ray group catalogue and AMICO candidates with significant X-ray flux estimate, we created a sample of 622 candidate groups and clusters with optical and X-ray properties up to $z=2$ and down to less than $10^{13} M_\odot$.
 \begin{figure}
   \centering
   \includegraphics[width=8cm]{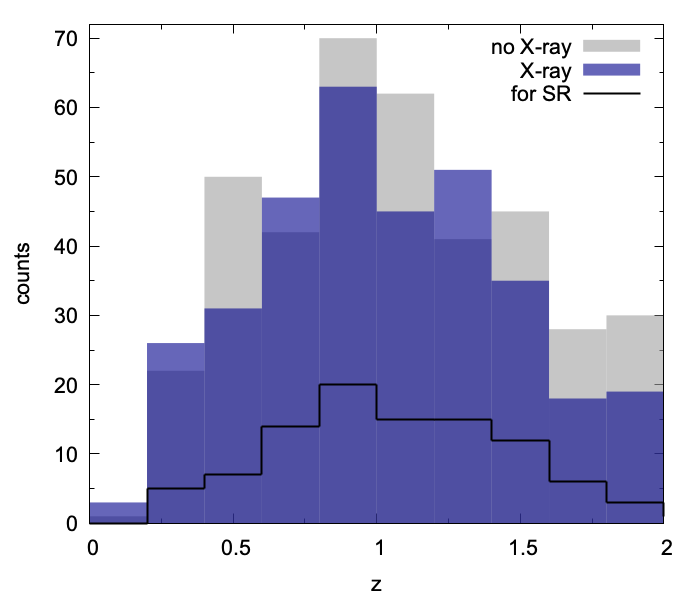}
      \caption{Distribution of the redshift of new detections analysed in the X-rays (from the $Y$-band run catalogue). 
      The grey histogram represents detections having X-ray flux below the 1$\sigma$ significance limit. Blue histogram refers instead to detections with significant X-ray emission ($\sigma>1$). The solid line highlights the subsample selected to study the relation between AMICO mass-proxies and X-ray mass.}
         \label{hist_xraynew}
   \end{figure}
\begin{figure*}
   \centering
   \includegraphics[width=18cm]{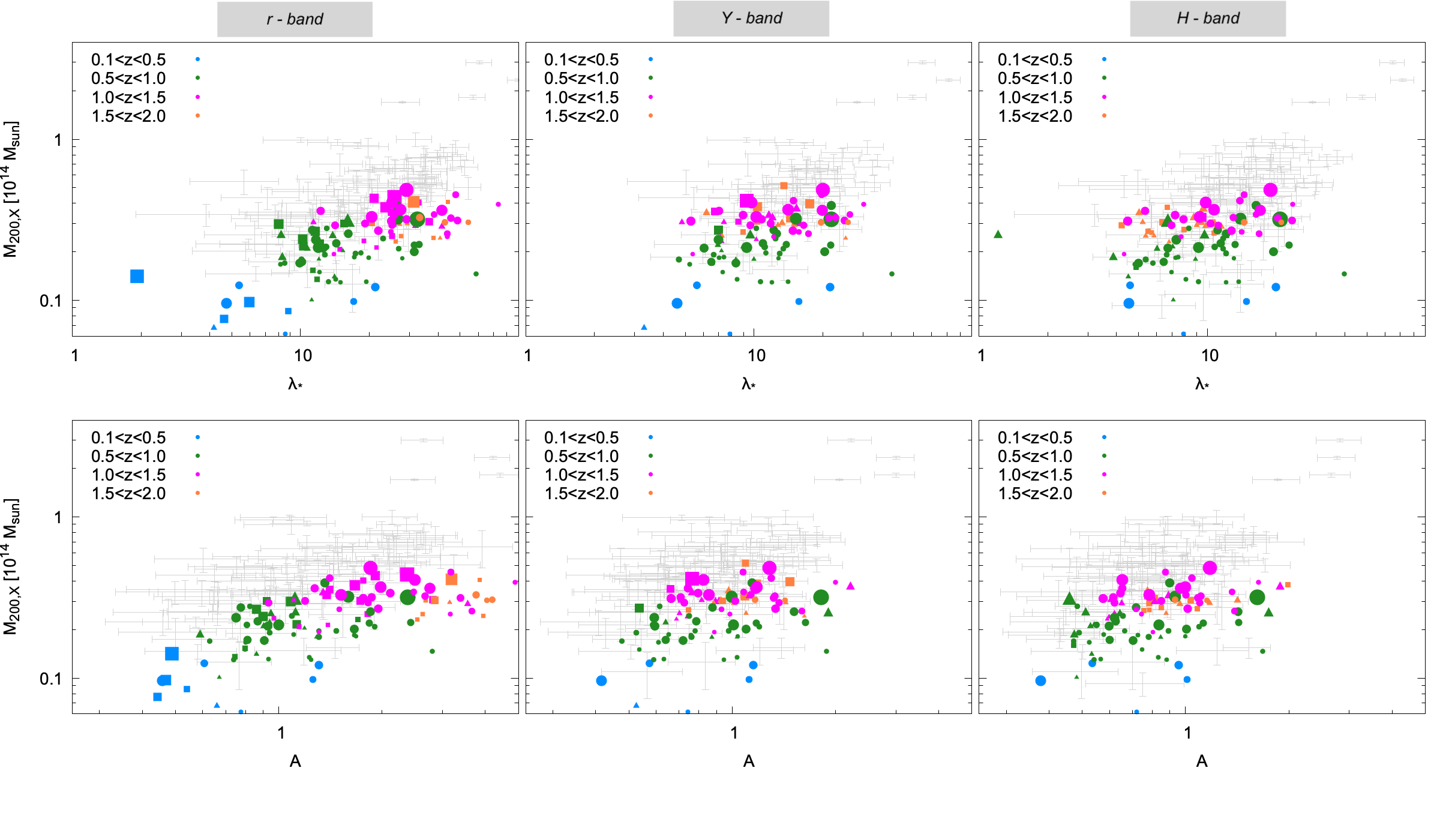}
      \caption{Same as in Fig. \ref{scalrelfig}, but for the sample of new detections with X-ray flux estimate selected for the scaling relation study (coloured points) in the full redshift range, $z \in $ [0.1, 2.0]. From left to right: detections from analysis in $r$, $Y$ and $H$-band. Different colours refer to different redshift bins, as labelled in the plot. Circles are detections found in all three runs, triangles are found in 2 out of 3 runs and squares indicate the new detections from respective runs. The size of the symbols is proportional to the X-ray flux significance. Grey error bars in the background indicate as a reference the detections used for the scaling relation derived from the direct matches with the X-ray catalogue (the same displayed in Fig. \ref{scalrelfig}).}
         \label{scalrelnewdet}
\end{figure*}
 
We repeated the calibration of the relation between AMICO mass proxies and X-ray mass, this time for the full sample of AMICO detections with X-ray estimates.
In order not to include biased estimates due to fore-/background or nearby contamination, we rejected all detections lying within 0.02 deg from the centre of another rich detection or lying on the extended X-ray emission of a bright cluster. This left us with $\sim 40\%$ of the full sample, which we used to perform scaling-relation calibration.
In Figure \ref{hist_xraynew} we show the distribution in redshift of the new detections analysed for the $Y$-band run catalogue, displaying the detections with and without (blue and grey) significant X-ray flux and the ones eventually selected for the study of the mass-observable scaling relations (black solid line).
In Figure \ref{scalrelnewdet}, we show the consistency between detections matched with the X-ray catalogue (shown by grey error bars) and new detections with X-ray estimates above significance limit (coloured points), used for the scaling-relation calibration. Most of the new detections lie within the scatter of the catalogue-matched detections, consistently with the trends found by the best-fitting scaling relations. We found that at fixed amplitude or $\lambda_\star$, new detections tend to be on average less luminous/massive when compared to successful matches with the catalogue. 
This is due to the lower X-ray $S/N$ level adopted in our analysis here and it shows the high completeness in the identification of bright X-ray extended sources of the previous X-ray catalogue \citep{gozaliasl19}.

\subsection{Redshift dependence of the scaling relations}

If we consider the whole sample of detections, including both successful matches with the X-ray catalogue and the new detections with X-ray flux estimates, we have the opportunity to study the scaling relations for clusters and groups extending up to $z \sim 2$ and over a wide range of mass/richness. This also allows for a study of the redshift-dependence of the scaling relations. We considered the scaling relation in Eq. \ref{bg_sr}, this time including the $z$-dependent term (i.e. $\gamma \neq 0$).

\begin{figure*}
   \centering
   \includegraphics[width=18cm]{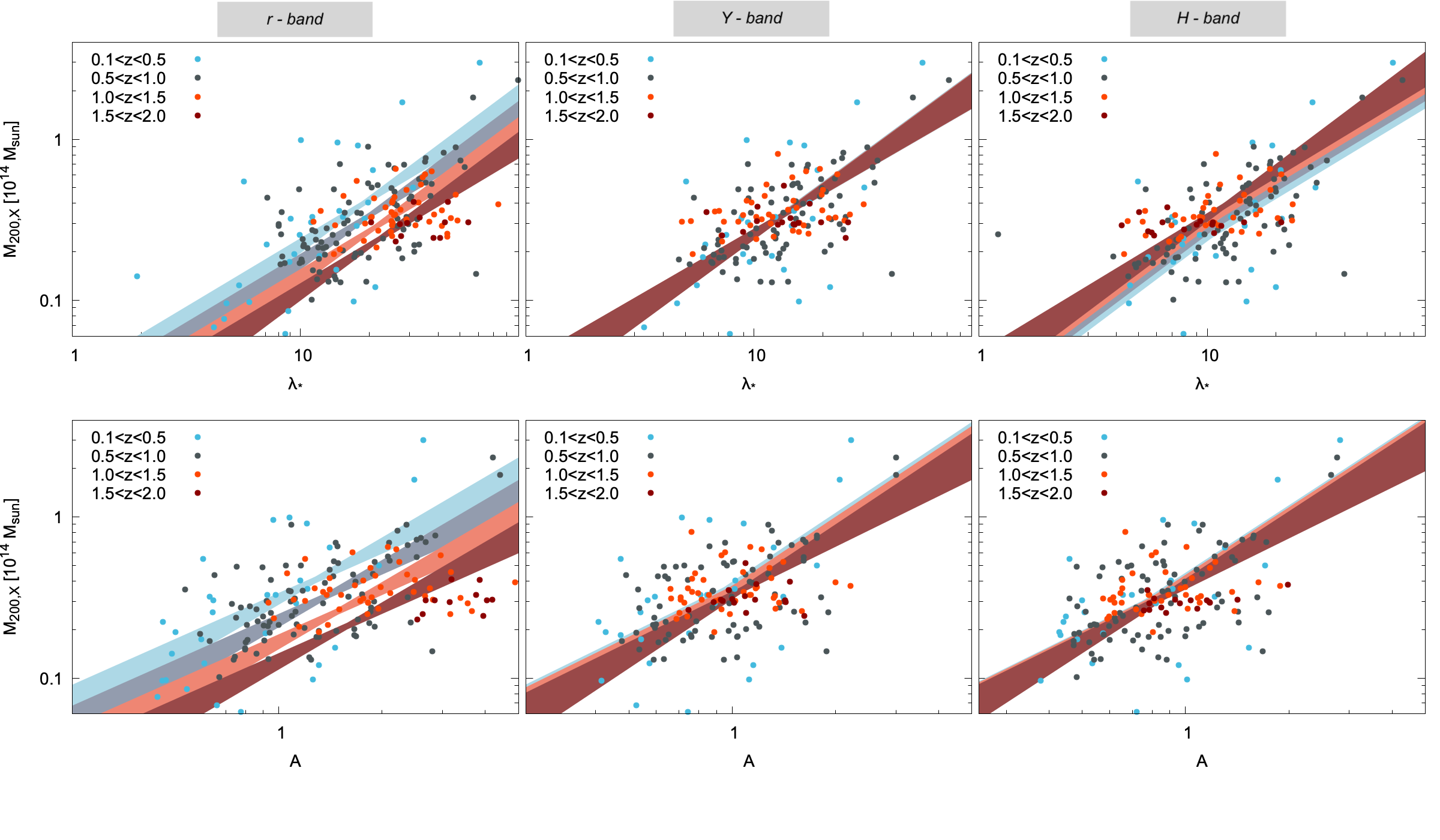}
      \caption{Relation between X-ray mass and richness (top panels) or amplitude (bottom panels) for the full sample including matched detections with the X-ray catalogue and new detections with X-ray flux estimate in the full redshift range, $z \in $ [0.1, 2.0]. From left to right: detections from analysis in $r$, $Y$ and $H$-band. Different colours refer to different redshift bins, as labelled in the plot. The redshift-dependent best-fitting relation is shown for typical redshift values representing the four redshift bins, $z=[0.4, 0.8, 1.2, 1.6]$ ($z$ increases from light blue to dark red) and with their 1$\sigma$-error region. Best-fit parameters are also reported in Table \ref{scalrel_zdep_table}. The redshift trend visible for $r$-band-based detections fades out for redder bands, where scaling relations become consistent with no redshift dependence.}
         \label{scalrel_zdep}
\end{figure*}

We chose $z_{ref}=0.9$, namely the median redshift of the entire studied sample. The results of the fitting analysis done with the term of redshift dependence, including the scatter standard deviation as well as the chosen pivot values, are shown in Table \ref{scalrel_zdep_table}. Best fitting relations are also shown in Figure \ref{scalrel_zdep} where the data-points are divided in four redshift bins with the relevant scaling relations computed at typical redshifts of the bins ($z=[0.4, 0.8, 1.2, 1.6]$).

We found a negative redshift dependence for the $r$-band detected clusters and groups, for which at fixed amplitude or $\lambda_\star$ more distant objects tend to have lower masses. This trend seems to fade out for redder bands, with an albeit negligible hint of an inverted trend for the relation $M_{200}-\lambda_\star$ in the $H$-band catalogue.

\begin{table*}
      \caption[]{Mass-proxy scaling relation parameters and pivot values (see Eq. \ref{bg_sr}) including redshift dependence based on the full selected sample with X-ray flux estimate.  The standard deviation of the scatter $\sigma$ is given in $\log_{10}$ space, being the scaling relations in the form $Y=\alpha +\beta X + \gamma Z \pm \sigma$, with $Y=\alpha +\beta X + \gamma Z$ referring to Eq. \ref{bg_sr}.}
         \label{scalrel_zdep_table}
     $$ 
         \begin{tabular}{c c c c c c c}
            \noalign{\smallskip}
            \textbf{\textit{r}-band} & & & & &  \\
            \hline
            \hline
             &   $\alpha$ & $\beta$ & $\gamma$& $\sigma$ & $O_{piv}$ & $z_{ref}$  \\
            \noalign{\smallskip}
            \hline
            \noalign{\smallskip}
            $\lambda_\star$ & $-0.498 \pm 0.019$ & $0.954 \pm 0.105$ & $-0.966 \pm 0.241$ & $0.170 \pm 0.016$ & 20 & \multirow{2}{2em}{0.9} \\
            \noalign{\smallskip}
             $A$  & $-0.516 \pm 0.021$ & $1.098 \pm 0.151$ & $-1.309 \pm 0.319$ & $0.182 \pm 0.017$& 1.4\\
            \noalign{\smallskip}
            \hline
            \hline
            & & & \\
            \noalign{\smallskip}
            \textbf{\textit{Y}-band} & & &  \\
            \hline
            \hline
             &   $\alpha$ & $\beta$ & $\gamma$ &$\sigma$ & $O_{piv}$ & $z_{ref}$  \\
            \noalign{\smallskip}
            \hline
            \noalign{\smallskip}
            $\lambda_\star$ & $-0.482 \pm 0.019$ & $0.931 \pm 0.110$ & $-0.010 \pm 0.172$ &$0.160 \pm 0.016$ & 13 & \multirow{2}{2em}{0.9} \\
            \noalign{\smallskip}
             $A$  & $-0.452 \pm 0.020$ & $1.232 \pm 0.187$ & $-0.206 \pm 0.194$ &$0.171 \pm 0.018$& 1.0\\
            \noalign{\smallskip}
            \hline
            \hline
            & & & \\
            \noalign{\smallskip}
            \textbf{\textit{H}-band} & & &  \\
            \hline
            \hline
            &   $\alpha$ & $\beta$ & $\gamma$ &$\sigma$ & $O_{piv}$ & $z_{ref}$  \\
            \noalign{\smallskip}
            \hline
            \noalign{\smallskip}
            $\lambda_\star$ & $-0.516 \pm 0.020$ & $0.962 \pm 0.103$ & $0.424 \pm 0.182$ &$ 0.145 \pm 0.017$& 11 & \multirow{2}{2em}{0.9} \\
            \noalign{\smallskip}
             $A$  & $-0.519 \pm 0.022$ & $1.220 \pm 0.170$ & $-0.094 \pm 0.194$ &$0.166 \pm 0.018$& 0.8\\
            \noalign{\smallskip}
            \hline
            \hline
         \end{tabular}
     $$ 
   \end{table*}

\subsection{New detections without X-ray counterpart}

\begin{figure}
   \centering
   \includegraphics[width=9cm]{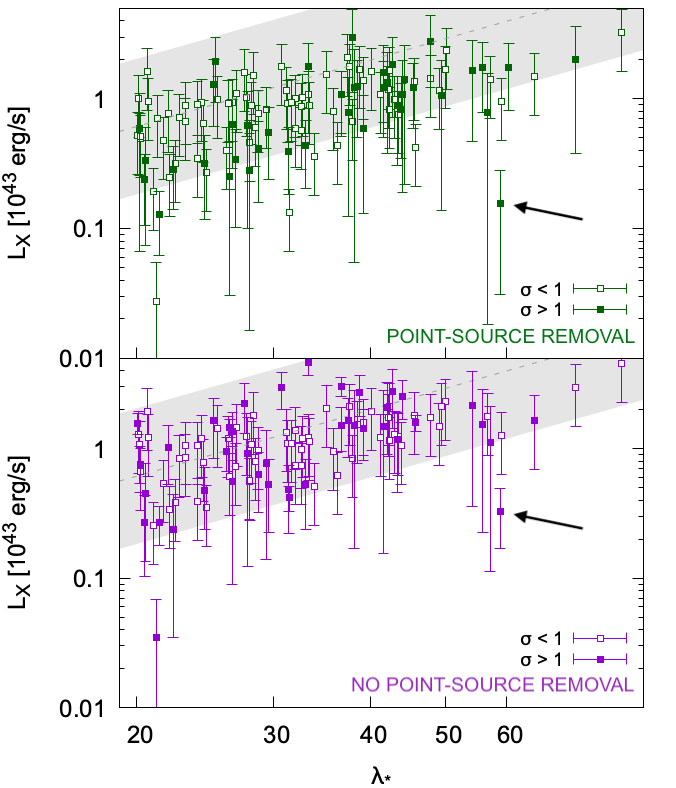}
      \caption{Relation between X-ray luminosity and intrinsic richness for the sample of richest detections ($\lambda_\star > 20$, $S/N >4.0$, $z<1.6$) that where found to have non-significant X-ray emission, with X-ray measurements performed with (top panel) and without (bottom panel) point-source removal. Empty squares are detections with X-ray flux below the $1\sigma$ significance limit. Grey shaded area in the background marks the $2\sigma$ scatter region relative to the $L_X-\lambda_\star$ relation found at $z_{ref}$ for the full $r$-band-detected sample. }
         \label{pointsources}
   \end{figure}
   
  \begin{figure}
   \centering
   \includegraphics[width=9cm]{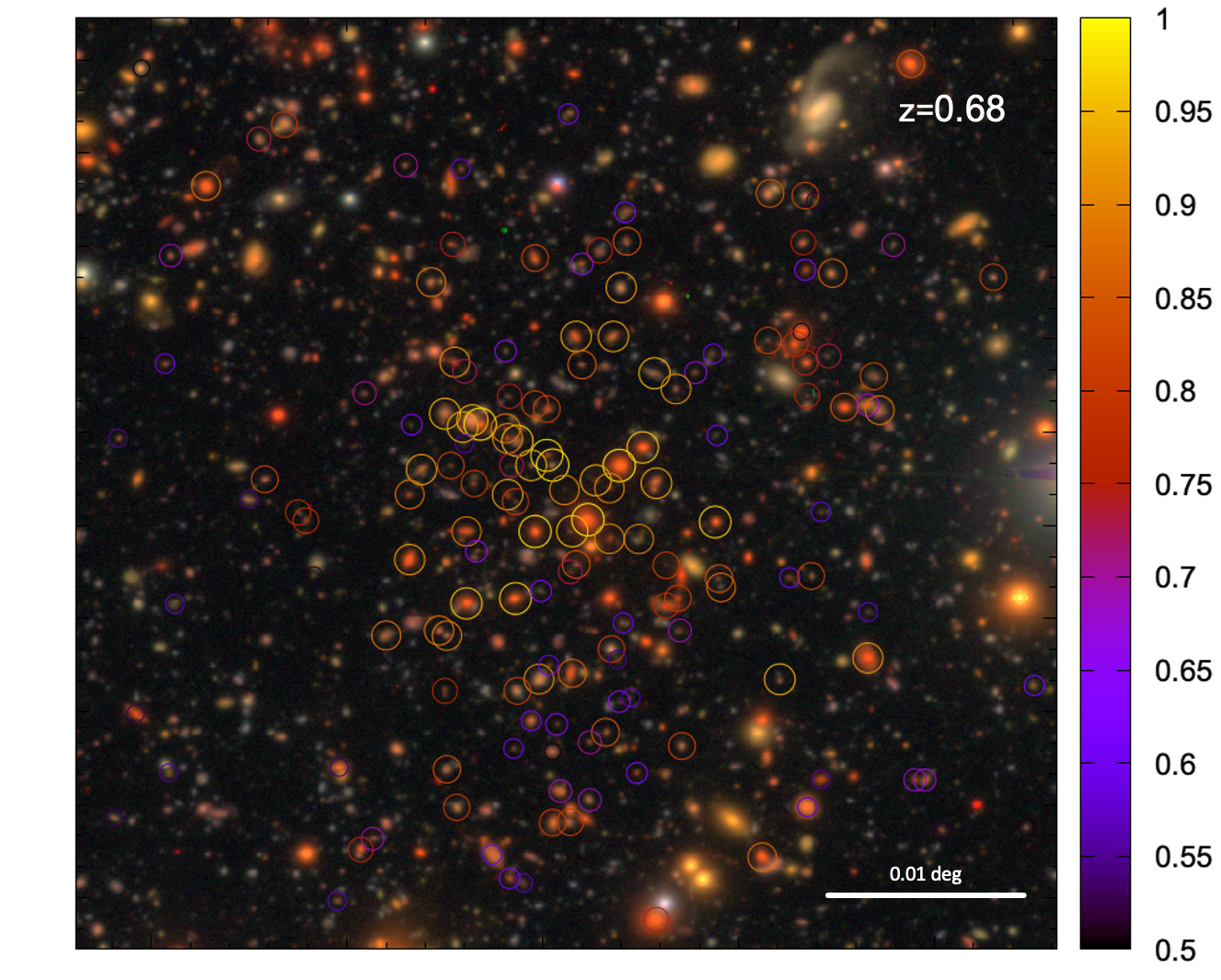}
      \caption{\textit{Detection 5}: a candidate cluster detected at $z=0.68$ with around 200 member galaxies ($\lambda_\star \sim 59$ in the $r$-band), without significant X-ray emission, even without point-source removal.}
         \label{examples_noxray}
   \end{figure}

As mentioned before, nearly half of the detections analysed in the X-rays were found not to have flux significance above unit. The relative number of new detections with and without significant X-ray flux does not seem to be significantly affected by cuts in $S/N$ or redshift (e.g. still $\sim 43\%$ of analysed new detections have $\sigma>1$ and $S/N >4.0$ at $z<1$). Among these detections we found and visually inspected several examples of clusters with a significant number of assigned galaxy members. In particular, we selected 116 groups with $\lambda_\star>20 $ (i.e. among our richest detections), detected with $S/N >4.0$ at $z<1.6$, which were found to have X-ray flux significance below $2\sigma$. We chose this threshold in order to include both detections below the significance limit and with low significance ($1<\sigma<2$).
We found a few cases of correspondence or vicinity to X-ray point sources. Thus, we investigated the impact of point-source removal in estimating flux and luminosity of extended source by repeating the measurements of X-ray properties for these 116 detections by considering the X-ray image without removal of point-sources. Undoubtedly, this represents a significant improvement for most of the analysed sources, as it is possible to see in Figure \ref{pointsources}, where we show the relation between $L_X$ and $\lambda_\star$ for this sample (compared to the scaling relation found for the full sample and the $2\sigma$ scatter) with (top panel) and without (bottom panel) point-source removal. Most of the selected detections become consistent with the rest of the sample by reintroducing the removed flux.
Nevertheless, this does not seem to be the case for two of the detections. These two outliers do not display significant improvement in their position with respect to the rest of the sample even with increased flux significance. One of these two, the lowest-richness object was found in all three runs at $z=0.36$, with $S/N \sim 5$, $\lambda_\star \sim 19$, $A \sim 1$, but with very low X-ray luminosity and large errors. The second object, was instead found at $z=0.68$ and for simplicity we refer to it as \textit{detection 5}. 

\subsubsection{\textit{The case of detection 5.}}
\textit{Detection 5} is one of the richest of the detections without X-ray flux significance above 2.0 and it has been detected with high signal-to-noise ratio in all AMICO runs ($S/N=6.6$). Its intrinsic richness, for instance, in the $r$-band analysis is $\lambda_\star \sim 59$. 
It was found at $z=0.68$, with $\sim 200$ galaxy members assigned with probability larger than $50\%$ (among these, almost half are assigned with probability $>75\%$). This is an example of detection whose X-ray properties measured without point-source removal did not improve sufficiently for it to become consistent with the rest of the sample in terms of the relation between optical and X-ray properties. \textit{Detection 5} is shown in Figure \ref{examples_noxray}, indicated by an arrow, with X-ray luminosity measured with or without point-source removal. 
The cluster redshift probability distribution, $P_{cl}(z)$, of \textit{detection 5} shows no anomalies and it is consistent with the $P_{cl}(z)$ of clusters at similar redshift, with similar richness which have been found also via X-ray selection. 

This interesting detection is compatible with being part of the COSMOS wall \citep{iovino16}. Using the same data used by \citet{abrilmelgarejo21}, retrieved with MUSE (\citeauthor{epinat21} \citeyear{epinat21}; Epinat et al., in prep.; \citeauthor{bacon10} \citeyear{bacon10}), we performed our dynamical analysis. The two groups mentioned in \citet{abrilmelgarejo21} (CG84 and CG84b) are responsible for the two peaks in the redshift distribution of AMICO members, when considering spectroscopic redshifts, centred at $z=0.6808$ (\textit{group 1}) and $z=0.6963$ (\textit{group 2}). Running the \texttt{Clean} algorithm \citep{mamon13} on these two peaks and considering $\pm 2000 \, km/s$ around each peak, results in the following characteristics for the two groups: \textit{group 1} has 19 clean spectroscopic members, velocity dispersion of $263 \pm 54 \, km/s$, and a corresponding radius, $R_{200}=397 \, kpc$; \textit{group 2} has 7 spectroscopic members, velocity dispersion of $170 \pm 62 \, km/s$, and $R_{200}=254 \, kpc$. We note that the mass estimates reported in \citet{abrilmelgarejo21} are associated with measuring higher values of velocity dispersion (of around $370 \, km/s$) compared to our estimate, which we attribute to the contamination of interlopers, which in our analysis are more efficiently rejected. The number of member galaxies reported in \citet{abrilmelgarejo21} is indeed much higher: 35 and 31 (compared to our 19 and 7). The presence of interlopers, possibly arranged in a filamentary structure elongated along the line of sight, is reflected in the large richness measured by AMICO, given by the presence of a large number of galaxies within the redshift range accessible to the photometric redshifts, and in the relatively small X-ray flux produced by this structure, which is compatible with our spectroscopic mass estimate.

\section{Conclusions}\label{conclusions}
We successfully detected galaxy clusters and groups in the range $0.1<z<2$ by applying the AMICO algorithm to a photometric galaxy sample mostly based on the COSMOS2020 catalogue. We did this in three independent runs by using magnitudes in three different bands as galaxy properties: $r$, $Y$ and $H$-band.
The final catalogue contains 1269 candidate clusters and groups among which 666 were detected with $S/N >3.5$ in at least one of the runs. Among the detected clusters, 490 were detected in all three runs. All three runs provided us with new detections.

The main achievements of this study can be summarised as follows:
   \begin{enumerate}
      \item we found that, with the creation of a suited cluster model, a composite mask and the proper regularisation of the noise model, the AMICO algorithm is efficient in detecting clusters also in peculiar survey configurations, such as the deep 2-deg$^2$ COSMOS field;
      \item we found 122 correspondences with the most recent X-ray selected group catalogue for the COSMOS field. Additionally, we confirmed X-ray emission with flux significance larger than 1.0 for other 500 AMICO detections, for a total sample of 622 AMICO candidate clusters and groups with associated X-ray properties up to $z=2$;
      \item the comparison with the X-rays allowed for the calibration of the scaling relations between AMICO mass-proxies and X-ray mass up to $z=2$ and down to less than $10^{13} \, M_\odot$. We found that the new detections with X-ray emission are consistent with the rest of the sample in terms of relation between X-ray and optical properties;
      \item the inclusion of the redshift dependence term in the scaling relation analysis showed that the $Y$-band and the $H$-band magnitudes used as galaxy properties for the cluster search result in a more stable relation between X-ray mass and AMICO richness/amplitude (w.r.t. the $r$-band magnitude), which is an important result for setting the redshift dependence of the calibration; 
      \item we investigated the impact of point-source removal on X-ray estimates and found that in many cases this can be source of underestimate of X-ray luminosity. We found nevertheless an interesting example of X-ray underluminous object with respect to its galaxy content as detected by AMICO. This is possibly due to the presence of interloper galaxies arranged in a filamentary structure aligned along the line of sight.
   \end{enumerate}
The creation of such a cluster catalogue for the deep COSMOS field, including robust mass proxies and lists of cluster members, paves the way to different kinds of studies. This cluster catalogue is a resource for the study of cluster galaxy population and galaxy evolution and for the study of the formation, evolution and physics of clusters themselves.

Moreover, this work represents a key step towards the successful exploitation of the AMICO algorithm in other peculiar survey configurations and proves the importance of the creation of suited input parameters.

\begin{acknowledgements}
      We gratefully acknowledge the contribution of the entire COSMOS collaboration, consisting of more than 100 scientists. More information on the COSMOS survey is available at \url{https://cosmos.astro.caltech.edu/}. We thank Micol Bolzonella for useful suggestions.
      LM acknowledges the support from the grants PRIN-MIUR 2022 20227RNLY3 and ASI n.2018-23-HH.0. GC acknowledges the support from the grant ASI n.2018-23-HH.
\end{acknowledgements}

\bibliographystyle{aa} % style aa.bst
\bibliography{references}

\end{document}